\documentstyle[12pt]{article}
\input epsf

\newcounter{fig}
\renewcommand{\thefig}{\arabic{fig}}
\def\newfig{\refstepcounter{fig} Fig.~\thefig.\ }

\def\bfig{\begin{center}}
\def\efig{\end{center}}

\def\A{{\cal A}}
\def\L{{\cal L}}
\def\O{{\cal O}}
\def\li2{{\rm Li}_2}
\def\alfa{\left({\alpha\over 2\pi}\right)}
\def\keV{{\rm keV}}
\def\MeV{{\rm MeV}}
\def\GeV{{\rm GeV}}
\def\TeV{{\rm TeV}}
\def\half{{\textstyle{1\over2}}}

\def\roughly#1{\,\,\raise.3ex\hbox{$#1$\kern-.75em\lower1ex\hbox{$\sim$}}\,\,}

\def\beq{\begin{equation}}
\def\eeq{\end{equation}}
\def\bea{\begin{eqnarray}}
\def\eea{\end{eqnarray}}

\def\Mpl{M_{\rm Planck}}
\def\vev{v}
\def\vevS{\langle S \rangle}
\def\vevFS{\langle F_S \rangle}
\def\sw{s_W^2}

\def\alphas{\alpha_3}
\def\alphat{\alpha_t}
\def\alphab{\alpha_b}
\def\alphatau{\alpha_{\tau}}
\def\lt{\lambda_t}
\def\lb{\lambda_b}
\def\ltau{\lambda_\tau}
\def\mt{m_t}
\def\mtpole{m_t^{\rm pole}}

\def\dmb{\delta m_b/m_b}

\def\dmtau{\delta m_\tau/m_\tau}
\def\meffA{m_{{\rm eff},A}}
\def\tb{\tan\beta}
\def\tbz{\tan\beta_{\widetilde \tau,\,\rm light}}
\def\tbc{\tan\beta_{\rm new\,vac}}
\def\tbt{\tan\beta_{\rm tunnel}}
\def\tbtbnd{\tan\beta_{\rm tun\,bnd}}
\def\SEb{S_E(\phi_b)}
\def\Dt{\Delta_t}
\def\Db{\Delta_b}
\def\Mess#1{M_{M_{#1}}}
\def\Mcrit{M_{M,{\rm crit}}}
\def\Mbar#1{{\overline{M}}_{#1}}
\def\Mhat#1{{\widehat{M}}_{#1}}

\def\mbar#1{{\overline{m}}_{#1}}

\def\alphbar#1{{\overline{\alpha}}_{#1}}
\def\lbart{{\overline{\lambda}}_t}

\def\Abart{{\overline{A}}_t}
\def\Abarb{{\overline{A}}_b}
\def\Abartau{{\overline{A}}_\tau}
\def\Abar{\overline A}
\def\Bbar{\overline B}
\def\mubar{\overline \mu}

\def\mstau{m_{\widetilde \tau}}
\def\mstauD{m_{\widetilde \tau,D}}
\def\mslep{m_{\widetilde L}}
\def\mslepD{m_{\widetilde L,D}}
\def\mlight{m_{\widetilde \tau,\,\rm light}}
\def\msq{m_{\widetilde Q}}
\def\mst{m_{\widetilde t}}
\def\msb{m_{\widetilde b}}
\def\msusy{\widetilde m}
\def\Lmin{L_{\rm min}}
\def\slep{\widetilde L}
\def\stau{\widetilde \tau}
\def\staumat{{\cal M}_{\stau,\slep}}
\def\epsmu{\epsilon_\mu}
\def\Vtt{V_{\tilde c \tilde t}}
\def\mtt{m_{\tilde c \tilde t}}

\begin{document}
\begin{titlepage}
\begin{center}
December 1996\hfill    UND-HEP-96-US01 \\
               \hfill    CERN-TH/96-349 \\
               \hfill    hep-ph/9612464
\vskip .2in
{\large \bf Large tan $\beta$ in  Gauge-Mediated SUSY-Breaking
Models}
\vskip .3in

\vskip .3in
Riccardo Rattazzi\footnote{E-mail:
rattazzi@mail.cern.ch}\\[.03in]
{\em Theory Division, CERN\\
     CH-1211 Geneva 23, Switzerland}
\vskip 10pt
Uri Sarid\footnote{E-mail:
sarid@particle.phys.nd.edu}\\[.03in]
{\em Department of Physics\\
     University of Notre Dame\\
     Notre Dame, IN 46556  USA}
\end{center}
\vskip .2in
\begin{abstract}
\medskip

We explore some topics in the phenomenology of gauge-mediated
SUSY-breaking scenarios having a large hierarchy of Higgs VEVs,
$v_U/v_D = \tan\beta \gg 1$. Some motivation for this scenario is
first presented. We then use a systematic, analytic expansion
(including some threshold corrections) to calculate the $\mu$
parameter needed for proper electroweak breaking and the radiative
corrections to the $B$ parameter, which fortuitously cancel at
leading order. If $B = 0$ at the messenger scale then $\tan\beta$ is
naturally large and calculable; we calculate it. We then confront
this prediction with classical and quantum vacuum stability
constraints arising from the Higgs-slepton potential, and indicate
the preferred values of the top quark mass and messenger scale(s).
The possibility of vacuum instability in a different direction yields
an upper bound on the messenger mass scale complementary to the
familiar bound from gravitino relic abundance. Next, we calculate the
rate for $b\to s\gamma$ and show the possibility of large deviations
(in the direction currently favored by experiment) from
standard-model and small $\tan\beta$ predictions. Finally, we discuss
the implications of these findings and their applicability to future,
broader and more detailed investigations.

\end{abstract}
\end{titlepage}

\section{Introduction}

Until recently, most phenomenological investigations of
supersymmetric (SUSY) models assumed that supersymmetry was broken in
a hidden sector, and that this breaking was communicated to the
visible sector, the minimal supersymmetric
standard model (MSSM), through interactions characterized by very
high scales, usually $\sim \Mpl$.  But in another class of models
supersymmetry is broken at a much lower scale, and communicated to
the standard model via gauge interactions. Dynamical supersymmetry
breaking (DSB) was typically associated with this gauge-mediated
SUSY-breaking (GMSB) scenario, since one could then explain the
hierarchy between the Planck and weak scales purely in terms of
field-theoretic dimensional transmutation. This second direction
received attention in the early days of SUSY model building
\cite{ref:old} (when it was sometimes called supersymmetric
technicolor), but was subsequently largely abandoned. The models were
very complicated and cumbersome, and little was known about the
dynamics of such strongly-coupled SUSY theories. Yet low-energy GMSB
models do offer an attractive feature: they dispose of the pressing
problem of flavor violation by soft SUSY-breaking terms. The flavor
problem is ubiquitous in gravity-mediated SUSY-breaking scenarios
(see, e.g., Ref~\cite{ref:kostel}), in string contexts
\cite{ref:kaplouis}, and more generally whenever the scale
$\Lambda_{\rm flavor}$ at which flavor physics arises is not far
above the scale $\Lambda_{\rm soft}$ at which the soft masses are
induced in the MSSM sector. In gauge-mediated scenarios the soft
terms are induced by the gauge interactions of the standard model at
a scale $\Lambda_{\rm soft}$ which can be much below $\Lambda_{\rm
flavor}$. Then the sfermion masses are essentially universal in
flavor space at the scale $\Lambda_{\rm soft}$, with the only source
of flavor violation relevant to low-energy phenomena being the Yukawa
matrices, precisely as in the standard model. This appealing solution
of the SUSY flavor problem has recently been revived by the promising
theoretical progress  in constructing realistic dynamical
SUSY-breaking models. A better understanding of strongly-coupled
supersymmetric theories (see Ref.~\cite{ref:prog} and references
therein) and how they may break SUSY dynamically was combined with
new and simpler ways to couple the SUSY-breaking sector to the
observable one \cite{ref:new}.  In this work we will accept the above
successes as motivation, but we will not be concerned directly with
the issue of dynamical SUSY breaking, and instead we focus on the
phenomenological consequences of its mediation to the observable
sector.

In the prototypical GMSB model, the breaking of supersymmetry appears
in the vacuum expectation value (VEV) of the $F$-term of a
standard-model-singlet field $S$ (possibly as a spurion parametrizing
SUSY breaking in some other fields), which couples to messenger
fields --- new fields which do carry standard-model gauge quantum
numbers and so communicate, via  the gauge interactions, the breaking
of supersymmetry to the MSSM sector. Such models are attractive for
several reasons. First, as stated above, because the soft
SUSY-breaking parameters are determined by gauge quantum numbers, not
only are the flavor and (possibly) the CP problems naturally solved,
but also the additional flavor-violating effects from superpartners
are readily calculable (as we exploit below in predicting the rate
for radiative bottom quark decay). For the same reason, the number of
free parameters is greatly reduced relative to the
gravitationally-mediated scenarios. Because the SUSY-breaking scale
can be much lower than the $\sim 10^{12}\,\GeV$
gravitationally-mediated SUSY-breaking scale, the gauge-mediated
scenario may be much more accessible to future experimental tests.
These models are also attractive because they offer rich,
well-defined and experimentally distinguishable alternatives to the
traditional phenomenology; indeed, there is already an excellent
candidate event in the latest CDF analysis \cite{ref:phenom}. But
there is another attraction: because the messenger scale may be quite
low, the renormalization-group evolution of the parameters is
qualitatively different from the gravitationally-coupled
hidden-sector models \cite{ref:babu}. Perhaps the most striking
difference is in the $B\mu$ parameter coupling the two Higgs doublets
in the scalar potential: if for some reason it vanishes or is
extremely small at the messenger scale, its generated value at the
weak scale is greatly suppressed relative to the other superpartner
masses (as first noted in Ref.~\cite{ref:babu}), both by the high
order at which it is generated \cite{ref:dnirs} {\it and} by a
completely fortuitous cancellation between the two major
contributions, which occurs when the messenger scale is as low as it
can be.

This last observation is the motivation for this investigation. We
calculate the 2- and 3-loop contributions to $B$, which allows a
prediction of $B\mu$ if we assume its tree-level value effectively
vanishes. Currently there are no compelling model-building arguments
in favor of, or against, the tree-level vanishing of $B$; indeed the
GMSB scenarios do not necessarily predict either $\mu$, the
superpotential coupling of the Higgs doublets, or $B\mu$, and usually
both have been treated as free parameters to be fixed
phenomenologically. But in the absence of unambiguous arguments one
way or another, we are guided by two rewards of the $B_{\rm tree} =
0$
assumption: first, as discussed in Ref.~\cite{ref:dnirs} and below,
the two CP-violating phases in the SUSY sector naturally vanish (thus
suppressing the electron's electric dipole moment, which is
generically unacceptably large for light sleptons); and second, a
hierarchy in the VEVs $v_U \gg v_D$ of the two Higgs doublets is
naturally induced by the small coupling between their scalar fields,
which accounts for the lightness of the bottom quark and tau lepton
even if their Yukawa couplings are comparable to the top-quark
coupling (such as would be predicted in a Yukawa-unified model). This
last feature contrasts sharply with the large-hierarchy scenario in
gravitationally-mediated models, in which the radiative contribution
to $B$ is as large as the other superpartner masses because of the
extensive RG evolution and because there is no fortuitous, natural
cancellation mechanism; then a large hierarchy, namely a small $B$,
must be {\it imposed} artificially on the model by tuning
\cite{ref:nelran,ref:hrs,ref:us} the GUT- or Planck-scale boundary
value of $B$ to cancel the radiatively-generated contribution to one
part in $\sim v_U/v_D$. In other words, in GMSB models one {\it
predicts} a top-bottom quark mass hierarchy when imposing, for
CP-related reasons for example, the boundary condition $B = 0$ at the
messenger scale, while in hidden-sector models such a hierarchy, if
it is to come from the Higgs sector, must be put in by hand through a
substantial fine-tuning of seemingly unrelated parameters.

With this in mind, we first introduce the minimal GMSB model, define
its parameters and establish some notation. We will confine all our
detailed investigations to this minimal model, although the
techniques and many of the qualitative results can easily be
generalized. We then present our assumptions, and some justifications
why they may be plausible or at least interesting. With these we
calculate $\mu$ and $B$ to obtain the VEV ratio $v_U/v_D = \tb$,
using a well-defined perturbative expansion which yields analytic
(rather than numerical, as in previous work) results, clarifies their
precision, and demonstrates what would be needed to improve this
precision. In passing, we also obtain a measure of the fine-tuning
required when the superpartners become heavy. Once the VEV hierarchy
is fixed, the bottom quark and tau lepton Yukawa couplings $\lb$ and
$\ltau$ may be fixed from the corresponding masses. But the large
couplings which result can destabilize the scalar potential, which
would generate VEVs for charged MSSM fields. We first calculate three
related bounds on $\ltau$ stemming from stability in the sleptonic
sector: an improvement on the familiar bound from the positivity of
the mass-squared in the ordinary charge-conserving vacuum, as well as
new bounds from the possibility of a true charge-breaking global
minimum and from the longevity of a false charge-conserving minimum.
We then use vacuum-stability arguments for $\lb$ to place upper
bounds on the messenger scale, which complement the cosmological
bound arising from an upper bound on the gravitino mass. Finally, we
study the prediction of the radiative bottom quark branching ratio
$b\to s\gamma$, and learn that it can depart dramatically from the
standard, 2-Higgs and low-$\tb$ models; in fact, the prediction of
the large $\tb$ model is favored by the current experimental
measurement of that branching ratio.

\section{The Minimal Model}

In the minimal gauge-mediated (MGM) model, an SU(2) doublet and an
SU(3) triplet messenger fields receive masses through their Yukawa
couplings to a single scalar field $S$. Their fermionic components
have masses $\Mess{i} = \lambda_i \vevS$ ($i = 2,3$) while their
scalar components are split about this mass by the expectation value
of the $F$ component of $S$: $m_i^2 = |\lambda_i \vevS|^2 \pm
|\lambda_i \vevFS|$. The breaking of supersymmetry is then
transmitted to the MSSM gauginos through 1-loop diagrams and to MSSM
scalar superpartners through 2-loop diagrams. We'll use an overbar to
denote quantities evaluated at the messenger scale $\Mess{}$ without
distinguishing in this notation between the various messenger masses.
The gaugino masses are
\beq
\Mbar{i} = {\alphbar{i}\over 4\pi} \Lambda g(x_i) \equiv \Mhat{i}
g(x_i);\qquad
i = 1,2,3\,,
\label{eq:gmasses}
\eeq
while the scalar masses are
\bea
\mbar{\widetilde{\alpha}}^2 &=&
2 \Lambda^2 \left[C_3 f(x_3) \left({\alphbar3\over 4\pi}\right)^2 +
                  C_2 f(x_2) \left({\alphbar2\over 4\pi}\right)^2 +
                  {3\over5} Y^2 f(x_1)
\left({\alphbar1\over 4\pi}\right)^2\right] \\
&=& 2 C_3 f(x_3) \Mhat3^2 +
    2 C_2 f(x_2) \Mhat2^2 +
    {6\over5} Y^2 f(x_1)
\Mhat1^2.
\label{eq:smasses}
\eea
These observable masses depend on the messenger and SUSY-breaking
sectors through the overall scale parameter
\beq
\Lambda \equiv \vevFS/\vevS.
\label{eq:lambda}
\eeq
and through the mass ratios
\beq
x_i = \Lambda/\Mess{i}, \qquad i = 2,3\,.
\label{eq:xi}
\eeq
We will often need the log of the ratio between a typical messenger
mass scale $\Mess{}$ and a typical superpartner mass, say the gluino
mass $M_3$:
\beq
L \equiv \ln \left({\Mess{}\over M_3}\right) \simeq \ln
\left({\Mess{}\over\Lambda}\right) +
\ln\left({4\pi\over\alphas}\right)
\ge \ln\left({4\pi\over\alphas}\right) \equiv \Lmin\,.
\label{eq:Ldef}
\eeq
The functions $g(x_{2,3})$ and $f(x_{2,3})$ are given in
Ref.~\cite{ref:martin}, while we have defined $g(x_1) \equiv \frac25
g(x_3) + \frac35 g(x_2)$ and similarly $f(x_1)$. The coefficient
$C_3$ equals 4/3 for squarks and 0 for sleptons, $C_2$ equals 3/4 for
SU(2) doublets and 0 for singlets, and the hypercharge is normalized
according to $Y = Q - T_3$. We will denote the SU(5)-normalized
hypercharge coupling by $\alpha_1 = g_1^2/4\pi = {5\over3} g_Y^2/4\pi
= {5\over3} \alpha_Y$ where $g_Y^2/g_2^2 = \sw/(1-\sw)$ and $\sw =
\sin^2\theta_W\simeq 0.23$. Notice that $x_i < 1$ is needed to insure
the messenger scalars have a positive mass-squared. For $x_i < 0.90$,
$0.96 < f(x_i) < 1.005$ so we will simply approximate $f(x_i) \simeq
1$. But since $g(x_i)$ can differ appreciably from unity when $x_i
\sim 1$, we will keep the dependence on $x_i$ in the gaugino masses.
For convenience, we've defined a reduced messenger-scale gaugino mass
$\Mhat{i} = (\alphbar{i}/4\pi) \Lambda$ which does not carry the
$g(x_i)$ factor, and therefore enters the scalar masses directly. The
reduced gaugino mass will serve as a more convenient mass scale than
the larger $\Lambda$, and we will often state our results in terms of
$\Mhat2$, which is within $2-3\%$ of the observable weak-scale SU(2)
gaugino (``wino'') mass $M_2$ when $g(x_2) \sim 1$, that is, when
$\Mess{2} \roughly{>} 2 \Lambda$.

To retain the successful gauge-unification predictions of the MSSM,
we will assume, as in most previous work, that the MGM model is
embedded in a grand-unified theory valid at very high scales;
indeed, that was the motivation for choosing the messengers to form a
complete SU(5) multiplet. Therefore we do not expect the messenger
Yukawa couplings $\lambda_{2,3}$ to differ very much. If $\lambda_2 =
\lambda_3$ at the GUT scale, we find at the messenger scale a ratio
$1 < \lambda_3/\lambda_2 \roughly{<} 1.5$, with the maximal value
obtained if the gauge couplings completely dominate the RG evolution.
In the following we will consider not only the expected case
$\Mess3/\Mess2 \sim 1.3$ but also, for comparison, the opposite case
$\Mess2/\Mess3 \sim 1.3$. Thus the minimal messenger and
SUSY-breaking sectors determine the MSSM masses in three ways: the
overall superpartner masses are $\msusy \sim {\alpha\over4\pi}
\Lambda$, the overall scale at which these masses are generated is
the messenger mass scale $\sim \Mess{} = \Lambda/x$, and there may or
may not be much sensitivity to the messenger splittings $x_2$ versus
$x_3$.

\section{$\mu$ and $B\mu$}

The MGM paradigm does not predict a definite value for the $\mu$
parameter which couples the up- and down-type Higgs doublets $H_U$
and $H_D$ in the superpotential. The SUSY-breaking bilinear scalar
coupling, $B\mu$, is also not unambiguously predicted: it may be
related to $\mu$ because both violate the Peccei-Quinn symmetry
between the up- and down type Higgs doublets $H_U$ and $H_D$, or it
may be related to the trilinear scalar couplings, the $A$ terms,
because both violate the same R symmetry. Indeed this R symmetry is
also violated by the gaugino masses, so an interesting feature of
gauge-mediated models is that gaugino masses and $A$ terms originate
from the same source ($\vevFS$), even though the latter arise at a
higher loop order. The result, as stressed for instance in
\cite{ref:dnirs}, is that the dangerous SUSY CP phase ${\rm
arg}(A^*M_3)$ vanishes naturally. We find it attractive to assume
that the same mechanism eliminates the other CP phase $\phi_B= {\rm
arg }(B^* M_3)$, in other words to assume that $B$, like $A$, is zero
(or negligibly small) at the messenger scale where SUSY-breaking MSSM
parameters arise ({\it i.e.} $\Bbar = \Abar = 0$).  Whether $B$ is
more like $A$ or more like $\mu$ is an open issue in gauge-mediated
SUSY-breaking models. Various preliminary proposals as to the origins
of $\mu$ and $B\mu$ have been made as of this writing
\cite{ref:new,ref:dnirs,ref:gdp,ref:gjp}, with a variety of results.
In any case, it seems at least a plausible and phenomenologically
attractive possibility to have the gauginos as the only source of $A$
and $B$, while $\mu$ is generated independently. We could imagine
that $\mu$ arises from an interaction $X H_U H_D$ with some field $X$
for which $\langle X \rangle \ne 0$ but $\langle F_X\rangle = 0$, for
example in an O'Raifertaigh-type model. Another option is to generate
$\mu$ in a completely different sector using different dynamics. This
is the route chosen in \cite{ref:dnirs}, where the $\mu$ term arises
from $1/M_P$ operators via an intermediate scale. In that situation
$\mu$ can be considered hard at the messenger scale, and $B$ turns
out to be zero at that scale.

In this work, we will treat $\mu$ as just an independent parameter
and not address its origin. We will focus on the case $\Bbar = 0$,
since as discussed above it offers an attractive solution to the SUSY
CP phase problem, and since it leads to an interesting large
hierarchy between the expectation values of the $H_U$ and $H_D$ Higgs
doublets. Our work could easily be extended to
the more general situation where $\Bbar$ is just very small, say
${\cal O}(\alpha)$.

Though we cannot predict $\mu$ from the microscopic physics, we can
calculate what value it must have to allow proper electroweak
symmetry breaking. The up- and down-type Higgs mass parameters
evaluated near the electroweak scale may be written as:
\bea
m_U^2 &=& m_H^2 + \mu^2 - \lt^2 \Dt^2 \nonumber \\
m_D^2 &=& m_H^2 + \mu^2 - \lb^2 \Db^2
\label{eq:mUmD}
\eea
where $m_H^2$ is the tree-level soft SUSY-breaking common Higgs mass
$\overline m_{H_U}^2 = \overline m_{H_D}^2$ plus the
approximately-common corrections induced by gaugino loops, while
$\lt^2 \Dt^2$ and $\lb^2 \Db^2$ are the splittings in the Higgs
masses due to Yukawa couplings. The latter are normalized such that
the tree-level top- and bottom-quark masses are $m_t = \lt \vev
\sin\beta$ and $m_b = \lb \vev \cos\beta$, where $\vev = 174\,\GeV$.
Since, as we will see below, electroweak symmetry breaking is
dominated by $\langle H_U\rangle$, {\it i.e.} $\tb\gg 1$, there is a
direct relation between the up-type Higgs mass and the Z mass. Using
the tree-level
Higgs potential, one finds $m_Z^2 = -2 m_U^2 = -2 m_H^2 - 2 \mu^2 + 2
\lt^2 \Dt^2$. However, since the top squarks are considerably heavier
than the weak scale, the effective quartic Higgs coupling is
appreciably affected by top and bottom quark loop corrections
\cite{ref:higgsmass}, so that a more correct relation is $m_Z^2
(1+\delta_H) = -2 m_U^2$, where
\beq
\delta_H = {3\over\pi^2} {\lt^4\over g_Y^2+g_2^2} \ln\left({\mst\over
m_t}\right)
\label{eq:delH}
\eeq
is typically $\sim 1$. Therefore a more complete MGM model would need
to generate the following value for the $\mu$ parameter:
\beq
\mu^2 = \lt^2 \Dt^2 - m_H^2 - \half m_Z^2(1+\delta_H)\,.
\label{eq:musol}
\eeq
We will next calculate the various contributions to the right-hand
side of Eq.~(\ref{eq:musol}). For our present purposes, this value
for $\mu$ will be useful in estimating how finely such MGM models may
need to be tuned, and in calculating other predictions of these
models such as $\tan\beta$, the existence of charge-breaking minima,
and the rate for $b\to s\gamma$. In the future, when $\mu$ is
predicted by some model, Eq.~(\ref{eq:musol}) could be used to check
whether electroweak symmetry is properly broken in such models, and
to determine how massive must the superpartners be relative to $m_Z$.

A comment on the effective field theory picture underlying all of our
analysis is in order. Since we expect the hierarchy $\msq^2 > \mu^2
\gg m_Z^2$ it makes sense to use the MSSM at scales above $\sim \msq$
and the 2-Higgs standard model plus higgsinos and weak gauginos
between $\msq$ and $\mu$. The second Higgs doublet $H_D$ will have a
mass $\sim \mu$, so it may sensibly be integrated out at that scale,
leaving behind the standard model (with some additional
weakly-coupled gauginos and sleptons) between $\sim\mu$ and $\sim
m_Z$. From this perspective the parameters determining the masses
$m_{U,D}^2$ in the above equations should be evaluated near the
squark mass scale. (Notice that we neglected terms other than
$\delta_H$ in the Higgs Lagrangian evolution below $\msq$, such as
wave function renormalization. This is because their effect is
smaller, and because they affect only the last term in
Eq.~(\ref{eq:musol}), which is already just a perturbation on $\mu$.)

We first need to calculate the Higgs mass splitting $\lt^2 \Dt^2$
generated by the top Yukawa coupling. In first approximation, $\Dt^2$
is obtained from the leading term in the RG evolution of $m_U^2$,
which is simply proportional to $L$. To be more precise, we should
keep the $L^2$ term in the solution to the 1-loop RGE, and also
calculate the 1-loop threshold corrections (which determine at what
scale to  start and stop the RG evolution). The former is
straightforward and is  presented below, but the latter requires a
more involved calculation,  the result of which we will call
$c_\Delta$ and leave unevaluated. (The calculation of  $c_\Delta$
amounts to computing the finite part of a set of 3-loop diagrams at
zero external momentum in the original theory with messengers.) We
thus find, to a good approximation,
\bea
\Dt^2 &\simeq& {3\over8\pi^2} (\mbar{\widetilde Q}^2 +
\mbar{\widetilde t}^2 + \mbar{H}^2) L_{\Delta}
\left[ 1 - \frac12
\left(\frac{16}{3} {\alphbar3\over 2\pi} + 3
{\alphbar2\over2\pi}\right)
L_{\Delta} \right]
\nonumber\\
& & \phantom{.} + {3\over8\pi^2}
\left(\frac{16}{3} {\alphbar3\over2\pi}\Mbar3^2 + 3
{\alphbar2\over2\pi}\Mbar2^2\right) L_{\Delta}^2 +
c_{\Delta}
\label{eq:Dtone}\\
&\simeq& {2\over\pi^2} \Mhat3^2 L_{\Delta}
\left[1 + {9\over16}\left(\alphbar2\over\alphbar3\right)^2 -
\frac12
\left(\frac{16}{3}{\alphbar3\over2\pi} - 2 {\alphbar3\over2\pi}
g^2(x_3) + 3
{\alphbar2\over2\pi}\right) L_{\Delta} \right] + c_{\Delta}
\label{eq:Dttwo}\\
&\simeq& {2\over\pi^2} \Mhat3^2 L_{\Delta} + c_{\Delta}
\simeq 1.0 \Mhat3^2 + c_{\Delta}\,.
\label{eq:Dtappr}
\eea
where we estimate the log to be $L_{\Delta} = \ln(\Mess3/\msq)$, and
in the last line we display first a useful analytic approximation
(since the higher-order terms roughly cancel), and then a
representative value using $\Mess3 = 2 \Lambda$, $\alphas(m_Z) =
0.12$, and $\Mbar2 = 150\,\GeV$. To calculate $\Mhat3 =
(\alphbar3/\alphbar2) \Mhat2$ we use the 2-Higgs standard model to
evolve $\alphas$ between $m_Z$ and $\msq$ and the MSSM to evolve it
to the scale $\Mess3$. We cannot know $c_\Delta$ without further
calculation. It determines the precise value of $L_\Delta$, and is
expected to change the log by $\sim 1$. Thus we will admit a relative
uncertainty in $\Dt^2$ of $\sim 1/L_\Delta \roughly{<} 20\%$, and
will consider below the impact of this possible correction on the
prediction of $\tb$. (For very heavy messengers, $L_\Delta$ is large
so $c_\Delta$ is less significant; in that case our expressions above
are still no more precise than $\sim 20\%$, but now because we have
not kept the $L_\Delta^3$ terms, which is easily ameliorated.)

We also need a value for the top-quark Yukawa coupling near the
squark mass scale. We will evolve the $m_Z$-scale value
\beq
\lt(m_Z) = {\mt(m_Z)\over \vev\sin\beta} \simeq
{\mtpole\over177\GeV}
\simeq {180\pm15\GeV\over177\GeV} \simeq 1.02 \pm 0.08
\label{eq:ltZ}
\eeq
up to the squark mass scale using the 2-Higgs standard model RG
equation (and thereby neglecting the change in the coefficient of
$\lt^2$ when passing from the 1-Higgs to the 2-Higgs standard models
at a scale $\sim\mu$):
\beq
\lt \equiv \lt(\msq) \simeq \lt(m_Z) \left[1 -
{4 \alphas - 3 \alphat\over 2\pi} \ln\left({\msq\over m_Z}\right)
\right],
\label{eq:lt}
\eeq
where $\alpha_t \equiv \lt^2/(4\pi)$ (and similarly
$\alpha_{b,\tau}$).

Finally, we need
\beq
m_H^2\simeq {3\over2} \Mhat2^2 \left[
1 + \frac15\left({\alphbar1\over\alphbar2}\right)^2 +
\left\{{2 \alphbar2 g^2(x_2) \over2\pi}\right\}  L\right]
\label{eq:mH}
\eeq
which approximates the common Higgs mass to within $\sim1\%$. By
inserting Eqs.~(\ref{eq:Dttwo}), (\ref{eq:ltZ}), (\ref{eq:lt}) and
(\ref{eq:mH}) into Eq.~(\ref{eq:musol}), we can determine the value
of the $\mu$ parameter needed for proper electroweak breaking. While
we will use the full expressions in predicting $\tb$, a good
approximation to $\mu$ is simply:
\beq
\mu^2 \simeq \lt^2 \left({2\over\pi^2}\right)\Mhat3^2 L - \frac32
\Mhat2^2 - m_Z^2
\label{eq:amu}
\eeq
accurate to $\sim5\%$ (ignoring $c_\Delta$). Alternatively, we find
that the value of $\mu$ is a linear function (in qualitative
agreement with Ref.~\cite{ref:dtw}) of the wino mass $M_2$ to within
a few GeV, for $M_2$ significantly above $m_Z$: $\mu \simeq 25\,\GeV
+ k M_2$, where $k$ is given in Table~1 for $\alphas = 0.12$ and for
various choices of the messenger and top quark masses.

\begin{table}

\centerline{
\begin{tabular}{|c||c|c|c|}
\hline
$(\Mess3 = 1.3 \Mess2)$ & \multicolumn{3}{c|}{$m_t$ in GeV} \\
\hline
$\Mess2/\Lambda$ & $165$ & $180$ & $195$ \\
\hline \hline
1 & 0.80 & 1.01 & 1.22 \\
2 & 1.13 & 1.40 & 1.67 \\
10 & 1.33 & 1.62 & 1.91 \\
100 & 1.46 & 1.76 & 2.06 \\
10000 & 1.57 & 1.88 & 2.20 \\
\hline \hline
\end{tabular}
\hskip 1cm
\begin{tabular}{|c||c|c|c|}
\hline
$(\Mess2 = 1.3 \Mess3)$ & \multicolumn{3}{c|}{$m_t$ in GeV} \\
\hline
$\Mess3/\Lambda$ & $165$ & $180$ & $195$ \\
\hline \hline
1 & 0.93 & 1.18 & 1.42 \\
2 & 1.09 & 1.37 & 1.64 \\
10 & 1.27 & 1.56 & 1.85 \\
100 & 1.41 & 1.71 & 2.01 \\
10000 & 1.53 & 1.85 & 2.17 \\
\hline \hline
\end{tabular}
}

\caption[Table 1]{The coefficient $k$ in the linear approximation
$\mu \simeq 25\,\GeV + k M_2$ (which holds to within a few GeV), as a
function of the messenger and top quark masses. In the table on the
left, the triplet messenger mass is $30\%$ higher than the doublet,
as expected from RG evolution; the splittings are reversed in the
table on the right.}
\end{table}

With the full expression for $\mu$ we can calculate the chargino mass
matrix. The lightest chargino must be heavier than $\sim 100\GeV$;
the actual bound is either slightly higher from CDF if one assumes
the SUSY-breaking scale is low enough to produce a gravitino inside
the detector, or slightly lower from LEP under more robust
assumptions, but in any case the very light chargino case will
usually be of little interest to us. A lower bound of $100\GeV$ in
the large $\tb$ scenario implies $\left(M_2^2 - {1\over4}
m_Z^2\right) \left(\mu^2 - {1\over4} m_Z^2\right) \roughly{>} 2 m_W^2
(100\,\GeV)^2$. Since $\mu$ hardly ever exceeds $\sim 2 M_2$, we find
$M_2 \roughly{>} 85\,\GeV$. That is not necessarily the strictest
bound on the superpartner mass scale. For example, the SU(2)-singlet
tau slepton has $\mstau^2 \simeq {6\over5} \Mhat1^2 + \sw m_Z^2$,
where the last term comes from D-term contributions to the scalar
potential and protects $\mstau$ from becoming lighter than $\half
m_Z$ even when $\Mhat1$ is very small. However, as was pointed out
before and as we study in some detail below, the mixing term with the
SU(2)-doublet tau slepton can become appreciable when $\tb$ is large,
and can lower $\mstau^2$ unacceptably. Indeed, we will not be
interested in $M_2$ below $\sim m_Z$ precisely because such a light
gaugino mass is completely ruled out for large $\tb$ by the tau
slepton mass constraint.

What are the conditions for proper electroweak breaking? The
condition for a non-zero Higgs VEV (for large $\tb$) reads $m_U^2 <
0$, which may be written using Eq.~(\ref{eq:musol}) as
\beq
0 < \half m_Z^2 (1 + \delta_H) \simeq m_Z^2 \simeq (\lt^2 \Dt^2 -
m_H^2) - \mu^2.
\label{eq:mZpos}
\eeq
So if the superpartners are quite heavy and $\lt^2 \Dt^2 - m_H^2 \gg
m_Z^2$, then $\mu^2$ must cancel $\lt^2 \Dt^2 - m_H^2$ quite
accurately --- to a precision of roughly $m_Z^2/(\lt^2 \Dt^2 - m_H^2)
\sim m_Z^2/\mu^2$. Therefore when $\Mhat2 \sim m_Z$ this is very
little tuning in MGM models.  But it becomes more serious with
increasing superpartner mass. One measure of how precisely $\mu^2$
must be adjusted is
\beq
{m_Z^2\over\mu^2} \equiv \epsmu\,,
\label{eq:epsmu}
\eeq
which we will use to quantify the strength of the
electroweak-breaking constraint on any fundamental theory which
attempts to predict $\mu$. This is essentially equivalent to the
well-known fine-tuning constraint \cite{ref:bg} which must be imposed
on the MSSM to obtain a light Z mass from heavy superpartners. (The
tuning in GMSB was also studied in Ref.~\cite{ref:ciastru}.) In
Figs.~\ref{fig:bigone} and \ref{fig:bigtwo} we plot $1/\epsmu$ as a
function of the wino mass $M_2$. When the top Yukawa coupling is
relatively small and the messengers are relatively light so $L \sim
\Lmin$, the radiative electroweak-breaking parameter $\lt^2 \Dt^2$ is
naturally well below $\Mhat3^2 \sim 4.6 \Mhat2^2$, and subtracting
$m_H^2 \sim \frac32 \Mhat2^2$ reduces the final result $\mu^2$ almost
to $M_2^2$. Then a wino mass of even $400\,\GeV$, and hence a gluino
mass of $\sim 1\,\TeV$ and a squark mass of $\sim 1.6\,\TeV$,
requires a $\mu^2$ tuning of only 1 part in 25. Compare this with
gravitationally-mediated SUSY-breaking scenarios, in which the log in
the electroweak-breaking corrections to $m_U^2$ is replaced by
$\log(\Mpl/\msusy)\sim 6 \Lmin$, so a heavy stop generates a much
larger (negative) correction, which requires a much larger and more
precisely-tuned $\mu$.

\section{Calculating $B$ and $\tb$}

If $\Bbar = 0$ at a scale $\Mess{}$ only a few orders of magnitude
above the superpartner masses $\msusy$, the low-energy value of $B$
induced by RG evolution between $\Mess{}$ and $\msusy$ is
significantly smaller than $\msusy$. Since $B$ couples the up- and
down-type Higgs scalars, when $m_U^2$ becomes negative and $H_U$
acquires a VEV $v_U$, the VEV induced for $H_D$ is much smaller, so
$\tb = v_U/v_D \gg 1$. The induced $v_D$ is determined, up to
corrections that are ${\cal O}(1/\tan\beta)$ with respect to the
leading result, by the terms in the scalar potential that are linear
and quadratic in $H_D$. Upon minimizing that potential, we find
\beq
{v_D\over v_U} = {1\over\tb} = {B\mu\over \meffA^2}
\label{eq:tanbeta}
\eeq
where the ``effective'' pseudoscalar mass $\meffA$ is given by
\beq
\meffA^2 = m_H^2 + \mu^2 - \lb^2 \Db^2 - \half (1 - \delta_H') m_Z^2
= \lt^2 \Dt^2 - \lb^2 \Db^2 - \half m_Z^2 (1 + \half\delta_H -
\half\delta_H')
\label{eq:mA}
\eeq
and $\Db^2 \simeq \Dt^2$ to a very good approximation. In these
expressions we have included the renormalization of the $H_U^2 H_D^2$
quartic coupling due to top-bottom loops below the squark mass scale:
\beq
\delta_H' = {3\over\pi^2} {\lt^2 \lb^2\over g_Y^2+g_2^2} \ln
\left({\mst\over m_t}\right).
\label{eq:delHp}
\eeq
And as in Eq.~(\ref{eq:musol}) we have neglected smaller corrections.
With $\mu$ and $\meffA^2$ in hand, all that is needed for predicting
$\tb$ is the value of $B$ at low energies, which requires a knowledge
of the messenger-scale initial conditions and of the radiative
corrections. As previously stated, we will focus our attention on the
case $\Bbar = \Abart = \Abarb = \Abartau = 0$ (to sufficiently high
accuracy).

Since the $B$ parameter vanishes at tree level, its finite low-energy
value is generated completely by radiative corrections.
Diagrammatically we can identify two classes of contributions to $B$.
In the first class one of the two external Higgs lines is attached to
a higgsino-gaugino vertex; Fig.~\ref{fig:higgsino} shows the leading
effect in this class. In the second class one of the two external
Higgs lines is attached to a sfermion-antisfermion vertex;
Fig.~\ref{fig:mixed} gives the dominant effect. Such diagrams are
proportional to Yukawa couplings squared, and can be interpreted as
contributions to $B$ via generation of an effective $A$-term.  Thus
$B = B^G + B^A$. To lowest order, the 2-loop diagram of
Fig.~\ref{fig:higgsino} (or the lowest-order solution to the 1-loop
RG equation for $B$) yields $B^G \simeq -3 (\alphbar2/2\pi) L
\Mbar2$, while the 3-loop diagram of Fig.~\ref{fig:mixed} (again the
lowest-order RG solution) yields $B^A \simeq 8 (\alphbar3/2\pi)
[(\alpha_t+\alpha_b)/2\pi]L^2 \Mbar3$.  Note that, while $B^A$
requires an extra loop, it is a strong-interaction effect relative to
the weak $B^G$, so the two contributions are numerically comparable.
Indeed, when the messengers are as light as they can be, namely
$\Mess{} \simeq \Lambda$ and $L \simeq \Lmin$, one finds $B_G \simeq
-0.075 \Mbar2$ and $B_A \simeq  +0.07 \Mbar2$ (using $\lb \simeq
0.61$ and typical values for the other parameters). Because of this
fortuitous cancellation, $B$ is even smaller than the naive
expectation $B \sim (\alphbar2/2\pi) L \Mbar2$, and must be
calculated quite carefully.

\bfig
\leavevmode
\epsfysize=2.5cm \epsfbox{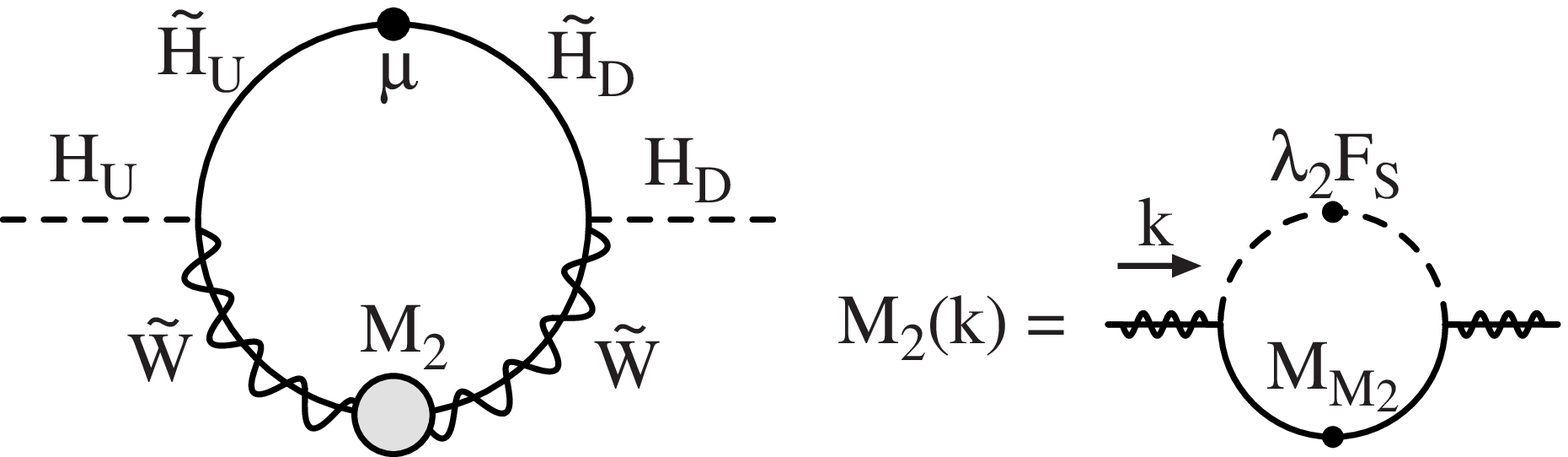}
\begin{quote}
{\small
\newfig \label{fig:higgsino} The leading contribution to $B^G$, which
is the direct gaugino-generated radiative correction to the $B$
parameter. The momentum-dependent wino mass $M_2(k)$ generated by the
messenger doublet loop is shown as a collapsed insertion on the left,
and as an expanded loop on the right.}
\end{quote}
\efig

\bfig
\leavevmode
\epsfysize=6cm \epsfbox{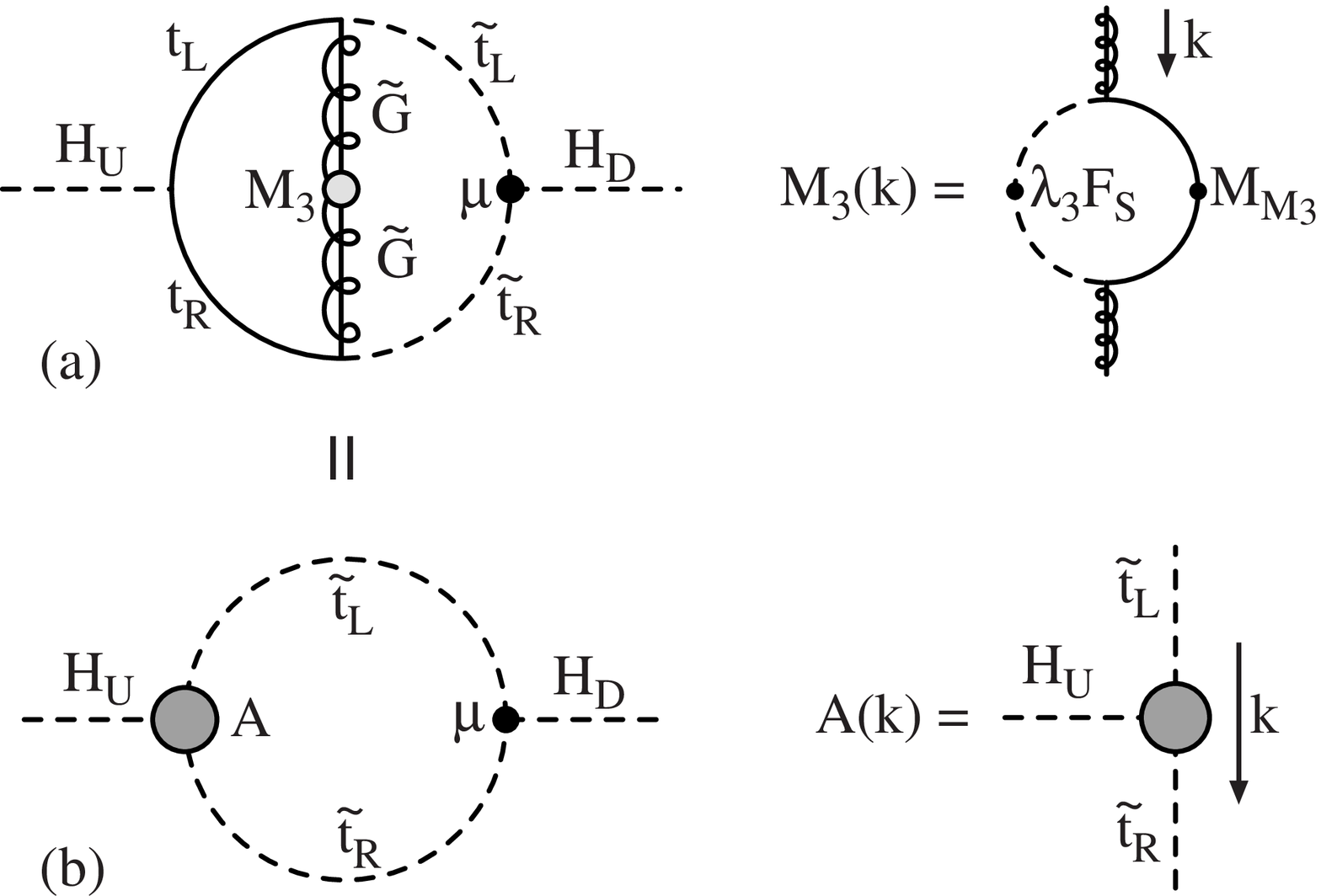}
\begin{quote}
{\small
\newfig \label{fig:mixed} The leading contribution to $B^A$, which is
the indirect gluino- and stop-generated radiative correction to the
$B$ parameter. In (a) the momentum-dependent gluino mass $M_3(k)$
generated by the messenger triplet loop is shown as a collapsed
insertion on the left, and as an expanded loop on the right. In (b)
the top-gluino loop has been collapsed into an effective
momentum-dependent $A(k)$ vertex.}
\end{quote}
\efig

In order to organize this calculation and keep track of the order to
which we are working, we'll write $B^G = B_1^G + B_2^G + B_3^G$ and
$B^A = B_1^A + B_2^A + B_3^A$, where each $B_i$ is generated in
proportion to the corresponding gaugino mass $M_i$, and further
expand each $B_i$ in powers of $L$ and coupling constants:
\beq
B_i^G= M_i {\alpha_i\over 2\pi}\left[ c_0 L + c_1 + c_2 \alfa L^2 +
c_3 \alfa L + \cdots \right]
\label{eq:BG}
\eeq
and
\beq
B_i^A = M_i {\alpha_i\over 2\pi} {\alpha_{t,b,\tau}\over 2 \pi}
\left[ d_0
L^2 + d_1 L + d_2 \alfa L^3 + d_3 +
d_4 \alfa L^2 + \cdots\right]
\label{eq:BA}
\eeq
where by $\alpha$ we mean one of $\alpha_{1,2,3}$ or
$\alpha_{t,b,\tau}$. The $\{c_k\}$ and $\{d_k\}$ are numerical
coefficients, different for each $B_i$, and which are expected to be
either of order 1 or zero. This second possibility occurs, for
example, in the lowest terms in $B_3^G$ since the leading
contribution arises only at 3-loop order and is  $\sim
\alpha_3\alpha_2^2 M_3 L$. When the messengers are not extremely
heavy, $L \sim \Lmin \sim 5$, and then $(\alpha/2\pi) L \sim 1/L \ll
1$ is a good classification, at least for the larger $\alpha$'s. Thus
we consider $(\alpha/2\pi) L$ and $1/L$ to be of the same order,
though we eventually drop the smaller $\alpha$'s when they are
numerically insignificant.

In the above, $\{c_0,d_0\}$ and $\{c_2,d_2\}$ can be interpreted as
the first two terms of an expansion in powers of $L$ of the solution
to the 1-loop RG equation for $B$. Then $c_1$ represents 1-loop
threshold effects at the messenger and superpartner scales. However
$d_1$ should then be viewed as arising from both 1-loop thresholds
and part of the 2-loop RG evolution: while in the 1-loop RG equation
for $B$ there are no direct contributions proportional to
$\alpha_{b,t,\tau} M_i$ (but only an indirect one via generation of
$A$-terms), the 2-loop RG equation does have such direct
contributions \cite{ref:mv}. We will not have to worry about the
origin of such terms; instead, we just calculate the diagram and keep
the appropriate terms. The higher-order terms $\{c_{k>2}\}$ arise
from higher powers of $L$ in the 1-loop RG solution and from 2-loop
(and higher) RG evolution and thresholds. The origin of $\{d_3\}$ is
just the finite 2-loop threshold at the messenger scale, essentially
the $L^0$ term in Fig.~\ref{fig:mixed}a. Again $d_{k>2}$ encompass
higher-loop RG effects, higher powers of $L$ in the 1-loop RG
solution and so on. We will only keep, and calculate, the terms
$c_{0,1,2}$ and $d_{0,1,2}$. To obtain the higher-order coefficients
would require substantial effort. For instance, to find the
$\{c_3,d_4\}$ terms to a consistent order, one needs not just the
2-loop  RG contribution but also the correspondingly accurate finite
contributions,  which include the radiative corrections to the
messenger-scale gaugino masses; since the latter are already a 1-loop
effect, their corrections require a 2-loop calculation. These effects
would require picking a renormalization scheme consistent with the
definition of the gauge and Yukawa couplings, since they can be
undone by $O(\alpha)$ changes in these couplings. We will later
discuss to what extent such higher-order effects could modify our
conclusions.

The contributions to $B^G$ and $B^A$ in our approximation arise from
the diagrams in Figs.~\ref{fig:higgsino} and \ref{fig:mixed}, and by
their 1-loop RG dressing which we do not display. The
momentum-dependent gaugino mass insertions on the gaugino lines are
just the messenger loop. We also define an effective,
momentum-dependent $A$ term $A(k)$ to be the loop in
Fig.~\ref{fig:mixed}a containing the third-generation fermions and
the gauginos, and represent this loop as an effective vertex in
Fig.~\ref{fig:mixed}b. To the order of interest, we do not need to
define a renormalization scheme and a regulator. The contributions to
$B$ are easily calculated by dividing the $dk$ integral, over
gaugino-higgsino momenta in Fig.~\ref{fig:higgsino} and over sfermion
momenta in Fig~\ref{fig:mixed}b, into three regions:
\beq
\int_0^{\infty} dk = \int_0^{k_{\rm IR}} dk + \int_{k_{\rm
IR}}^{k_{\rm UV}} dk + \int_{k_{\rm UV}}^\infty dk = I_{\rm IR} +
I_{\rm RG} + I_{\rm UV}
\label{eq:integrals}
\eeq
where $k_{\rm IR}$ and $k_{\rm UV}$ are near the superpartner masses
and the messenger masses, respectively. The integrals $I_{\rm IR}$
and $I_{\rm UV}$ then represent the low- and high-energy thresholds,
and contribute to $c_1$ and $d_1$ above. The bulk of the integration,
$I_{\rm RG}$, is just the lowest-order term in the RG evolution of
$B$ between $k_{\rm IR}$ and $k_{\rm UV}$, which we improve by adding
the next-order term using the RG equations; then $I_{\rm RG}$
contributes to $c_{0,2}$ and to $d_{0,1,2}$.

Rather than stating our results by giving the expressions for $c_k$
and $d_k$, we will absorb the threshold corrections $c_1$ and $d_1$
into the logarithmic terms by defining appropriate effective logs,
one for each contribution. In Eq.~(\ref{eq:BG}) we redefine $L \to L
+ c_1/c_0$ and in Eq.~(\ref{eq:BA}) we take $L \to L + \half
d_1/d_0$, thereby shifting $L \to L + {\cal O}(1)$; in the language
of RG evolution, this defines the correct scales at which to start
and stop the lowest-order RG logarithm. The other effects of this
redefinition are all contained in, or may be absorbed in, the
higher-order terms which we have already neglected. Note, for
example, that in our approximation order we cannot determine
precisely the higher-order RG logarithm, but we also don't need such
precision in this approximation. For simplicity we will use the same
log in the leading and next-to-leading RG evolution.

The calculation of $B^G$ is straightforward. In our approximation
$B^G = B_1^G + B_2^G$. The leading contribution to $I_{\rm RG}$ is
just the 1-loop RG evolution of $B$ to linear order in $L$, with zero
Yukawa couplings. The next-to-leading contribution requires adding
the next, quadratic order in $L$, and accounting for the thresholds
$I_{IR},I_{UV}$ by using effective values $L_{G1}$ and $L_{G2}$ for
the logs in the $M_1$ and $M_2$ contributions to $I_{\rm RG}$. We
find
\beq
L_{G2}=\ln\left({\Mess2 \over \mu}\right) + \left[1 + T(x_2)\right] -
{M_2^2\over \mu^2 - M_2^2} \ln\left({\mu \over M_2}\right)\,.
\label{eq:logweak}
\eeq
The addition of $1 + T(x_2)$ encapsulates the information about the
messenger-scale thresholds, while the last term properly accounts for
low-scale thresholds. The effective log $L_{G1}$ is similarly
defined, but we relegate its exact expression to the Appendix since
it involves a combination of doublet and triplet messenger masses and
is numerically less important. The threshold function $T(x)$ (also
defined in the Appendix) is bounded by $- 0.21 < T(x) < 0$, and
approaches $0$ very quickly as $x\to 0$. So we will neglect it,
setting $T=0$ in our calculation and thereby using $e \Mess2 \sim 2.7
\Mess2$ as the correct high-energy scale at which to begin the RG
evolution in $B^G$.

The calculation of $B^A$ is slightly more involved. We write $B_i^A
\sim \int A_i(k) dk/k$ as shown in Fig.~\ref{fig:mixed}b. Since the
$A$ terms, like the gaugino masses, are generated at the messenger
scale, they are suppressed above that scale by an inverse power of
the momentum $k$, and hence do not contribute a log to the $B$
integral $I_{\rm UV}^A$; their only contributions, to $d_3$ and
higher-order terms, are negligible for our purposes. For $k <
\Mess{}$, we find from Fig.~\ref{fig:mixed}b
\beq
A_{2,3}(k) = a_{2,3} \left({\alpha_{2,3}\over 2\pi}\right) M_{2,3}
\left\{\ln\left(\Mess{2,3}\over k\right) + \left[{3\over2} +
T\left(x_{2,3}\right)\right] + {\cal O}\left({k^2\over
\Mess{2,3}^2}\right) \right\}
\label{eq:Atwothree}
\eeq
where $a_2 = 3$, $a_3 = 16/3$ are the coefficients in the 1-loop RGE
for $A$. For $A_1(k)$, as before, we get an expression involving both
doublet and triplet messenger masses. We integrate $A_i(k)$ to obtain
$I_{\rm RG}^A$ and $I_{\rm IR}^A$. The $I_{\rm RG}^A$ integral will
have terms linear and quadratic in $L$, from the log and the constant
terms in $A(k)$, respectively, and these lead to $d_0$ and part of
$d_1$. The rest of $d_1$ comes from $I_{\rm IR}^A$ and is determined
by the sfermion masses. All of $d_1$ will be accounted for with
effective logarithms. Our results for the squark contributions to
$B^A$ are
\beq
L_{A2,3Q} = \ln\left({\Mess{2,3} \over \msq}\right) + 1\,,
\label{eq:logcol}
\eeq
while for the stau contributions they read
\beq
L_{A2L} = \ln\left({\Mess2 \over \mslep}\right) + \frac32\,,
\label{eq:logstau}
\eeq
where we have used the near degeneracy of the squarks but the
hierarchy $\mslep^2 \gg \mstau^2$ of the sleptons. Again we have
neglected $T(x)$ in the ultraviolet thresholds. Effective logs for
$B_1^A$ are also far too small to matter. Finally, the
next-to-leading RG contribution will be added on, yielding $d_2$ and
completing the calculation of $B^A$.

The above equations for the low-energy thresholds are very accurate
only when $m_Z^2 \ll \mu^2, \msq^2$. While the squarks are always
safely in this range, it is possible to choose parameters such that
$\mu\sim 200 - 300\,\GeV$. Then the full 1-loop effective potential
including the chargino and neutralino mass matrices would be more
accurate, but we do not expect this accuracy would be needed: for
example, the $c_1$ term in Eq.~(\ref{eq:BG}) would change by ${\cal
O}(m_Z^2/\mu^2)$, which is always a small correction to this
highest-order term in our expression, since inevitably $\mu$ is $2-3$
times larger than $m_Z$.

As an alternative to our approach of directly calculating the
diagrams in the full theory, one could have done the calculation in
the MSSM without messengers, using dimensional reduction (DR), and
matched to the full theory in some renormalization scheme. We have
checked that the two approaches give the same result. As already
noted above, the matching approach requires the inclusion of one term
in the 2-loop RG evolution, in order to obtain a result correct to
the given perturbative order; this term would partially furnish the
DR counterpart to the constant piece in $A(k)$.

With effective logs accounting for the threshold corrections, we can
obtain $B$ by solving the 1-loop RG equations to quadratic order in
$L$ and substituting the proper effective log for each $L$:
\bea
B &\simeq&
8 \Mbar3 \left({\alphbar3\over2\pi}\right)
         \left({\alphat + \alphab\over2\pi}\right)
         \left[L_{A3,Q}^2 +
             \left({\frac23\alphbar3 - \alphbar2\over2\pi}\right)
L_{A3,Q}^3
         \right] \nonumber\\
& & +
\phantom{8}\Mbar2 \left({\alphbar2\over2\pi}\right)
      \left[\frac92
\left({\alphat + \alphab
                          \over2\pi}\right) L_{A2,Q}^2 + \frac32
\left({\alphatau
                          \over2\pi}\right) L_{A2,L}^2 - 8
\left({\alphbar3\over2\pi}\right)
\left({\alphat + \alphab\over2\pi}\right) L_{A2,Q}^3
      \right] \nonumber\\
& & -
3 \Mbar2 \left({\alphbar2\over2\pi}\right)
      \left[L_{G2} - \left({\alphbar2\over2\pi}\right) L_{G2}^2
      \right] \nonumber\\
& & -
\frac35 \Mbar1 \left({\alphbar1\over2\pi}\right) L_{G2}
\label{eq:B}
\eea
In this expression we have dropped most of those terms which, though
formally of the same order as those we have kept, in fact contribute
less than 0.5\% of $B^G$ or $B^A$ and hence are numerically
insignificant (at least when $L$ is not many times larger than
$\Lmin$). Also, notice that we have chosen to express $B$ in terms of
the gauge couplings evaluated at the messenger scale but the Yukawa
couplings evaluated at the superpartner scale $\sim\msq$, using the
1-loop RG equations to relate the various scales. This choice turns
out to reduce the number of terms in the above expression.

We stress that we have now a complete, analytic expression for $B$ to
a consistent, well-defined perturbative order. At this order the $B$
we have calculated is scheme-independent, as is clear from
Eqs.~(\ref{eq:BG})-(\ref{eq:BA}). Our method does not require
regularization, nor any elimination of artificial scale dependences.
And no numerical integration was needed, nor could it improve our
result. Any further precision would require choosing a
renormalization scheme and evaluating higher-order effects such as
2-loop contributions to gaugino masses.

While we have carefully calculated $B$, the actual mass-squared
parameter in the scalar potential is $B\mu$ (indeed the diagrams of
\ref{fig:higgsino} and \ref{fig:mixed} all contain $\mu$). There are
of course radiative corrections to $\mu$, just as there were to $B$.
Those have not been included since they would only serve to relate
the messenger-scale value $\mubar$ to its low-energy value $\mu$. In
contrast to $B$, whose messenger-scale value we presume to know
($\Bbar = 0$), we place no such restrictions on $\mubar$, so we can
simply replace the unknown $\mubar$ and its multiplicative
corrections by $\mu$ at the superpartner scale with no loss in
predictivity. However, to the extent that $\mu$ is significantly
below the squark mass $\msq$, then between these two mass scales the
effective theory is not supersymmetric, and the $\mu$ appearing in
$B\mu$ and the $\mu$ appearing in the Higgs masses are radiatively
split. Moreover, the $\mu$ in $B^G\mu$ and the $\mu$ in $B^A\mu$
differ by ${\cal O}(\alpha_2/2\pi)$. Therefore Eq.~(\ref{eq:tanbeta})
for $\tb$ should be slightly corrected. But we expect these
correction to be no larger than the higher-order terms we have
explicitly dropped, and whose effects we will take into account in
our final results.

We have presented a careful calculation of the bilinear scalar
coupling $B$ in order to calculate $\tb$. But in the process we
needed to calculate the $A$ parameters, albeit to lower order, which
is useful for other investigations. We find for $A_t$ ($A_b$):
\beq
A_{t(b)} \simeq \frac{16}{3} \Mbar3 \left({\alphbar3\over2\pi}\right)
\left[L_3' + \left({3\alphbar3 - 3\alpha_{t(b)} - \half\alpha_{b(t)}
\over2\pi}\right) L^2\right] + 3 \Mbar2
\left({\alphbar2\over2\pi}\right)
\left[L_2' + \left({- 3\alpha_{t(b)}\over2\pi}\right) L^2\right]
\label{eq:Atb}
\eeq
where we have dropped all terms smaller than $\sim 1\%$ of $A_t$ (or
$A_b$), and
\beq
A_{\tau} \simeq -8 \Mbar3 \left({\alphbar3\over2\pi}\right)
\left({\alphab\over2\pi}\right) L^2 +
3 \Mbar2 \left({\alphbar2\over2\pi}\right) L_2' +
\frac95 \Mbar1 \left({\alphbar1\over2\pi}\right) L_1'
\label{eq:Atau}
\eeq
where we've dropped terms smaller than $\sim 5\%$ of $A_\tau$. The
effective logarithms are
\beq
L_{2,3}' = \ln\left({\Mess{2,3} \over M_{2,3}}\right) + 1 +
T(x_{2,3})
\eeq
while $L_1'$ is given by Eq.~(\ref{eq:Lone}) in the Appendix but with
the replacements $\mu \to M_1$ and $M_1 \to \mstau$. {\it Note} that
the above expressions for the $A$ terms correspond to zero external
momentum. The relevant expressions for processes involving on-shell
particles (such as corrections to sfermion masses or to two-body
decays) are easily obtained from the above by adding the appropriate
finite thresholds. We will use thye lowest-order expression for $A_t$
in Eq.~(\ref{eq:Atb}) for estimating the rate for $b\to s\gamma$.

At this point we should fix our sign conventions. We write the
superpotential as $W = \O_\mu + \sum_i \O_i$, where $\O_\mu$ is the
$\mu$ term and $\O_{t,b,\tau}$ are the Yukawa coupling terms. We fix
as the Yukawa Lagrangian $\L_Y= W$ where each pair of superfields is
replaced by the corresponding fermions. The Lagrangians for $A$- and
$B$-terms and gaugino masses are respectively $\L_{AB} = -V_{AB} =
B\O_\mu + \sum_i A_i \O_i$ and $\L_{1/2} = -M_i \tilde \lambda_i
\tilde \lambda_i$. With these conventions the radiatively induced
contributions to $B$ are $B^G\propto + M_2$ and $B^A\propto -M_3$, or
 ${\rm sgn}(B M_i) = |B^G| - |B^A|$, and after electroweak breaking
$v_d/v_u^* = +B\mu/m_A^2$. The crucial sign correlation is ${\rm
sgn}(\dmb) = - {\rm sgn}(B M_3 \mu \mu^*) = {\rm sgn}(-|B^G| +
|B^A|)$. To summarize: When the $A$-term contribution dominates $B$,
which we {\it define} to yield a positive $\tb$,  we also find a
positive correction to the bottom quark mass (that is, the $\lb$
defined in the MSSM must be increased by $\dmb$ before matching to
the observed $m_b$ in the standard model).

With a small $B$ first generated at 2-loop order, $\tb$ is large and
$\lb$ may become important; with a leading-order cancellation in $B$,
$\lb$ is further increased, and moreover the sensitivity of $B$ to
subleading-order effects is also increased. We must evaluate $\lb$
carefully, taking into account the standard-model (QCD) evolution of
$m_b$ between $m_b$ and $m_Z$, the subsequent 1-Higgs standard model
evolution up to the squark mass, and, most important, the large
finite threshold corrections $\dmb$ induced when matching to the MSSM
\cite{ref:hrs,ref:hemp}. From Ref.~\cite{ref:hrs} we find to a
sufficient approximation:
\beq
\lb \equiv \lb^{\rm MSSM}(\msq) \simeq {\eta_b^{-1} m_b(m_b)\over
174\,\GeV} \tb \left[1 -
{4 \alphas - \frac34 \alphat\over 2\pi} \ln\left({\msq\over
m_Z}\right) \right]
{1\over 1+\dmb}
\label{eq:lb}
\eeq
where $\eta_b = 1.48 + 0.08 \left[\alphas(m_Z)-0.12\right]/.01$ and
the running $b$ quark mass $m_b(m_b)$ is probably between $4.0$ and
$4.4\,\GeV$. The finite $\tb$ - enhanced threshold corrections are
given to within 10\% by
\beq
{\delta m_b\over m_b} \simeq \tb
\left[{\alpha_2(\msq)\over4\pi}\right] \left({\mu\over\Mhat2}\right)
\left[{\alpha_3(\msq)\over\alphbar3}\right]^2 \left[{g(x_2) \over
2}\right]
\label{eq:dmb}
\eeq
The latter can easily change $\lb$ by $\sim20\%$ and hence $\alpha_b$
by $\sim40\%$ and are crucial to a correct $\tb$ prediction, though
they have usually been omitted in previous work on this subject.

Inserting these various expressions into Eq.~(\ref{eq:tanbeta})
yields an equation for $\tb$ which may be solved self-consistently.
The thick curves in Figs.~\ref{fig:bigone} and \ref{fig:bigtwo} are
the predictions for $\tb$ as a function of $M_2$ for various values
of $m_t$, $\Mess2/\Lambda$ and $\Mess3/\Lambda$, with $c_\Delta = 0$.
We have not varied $\alpha_s(m_Z) = 0.12$ in these figures; when
varied by $\pm 0.005$, it typically affects $\tb$ by less than $\sim
\pm 1$, although for a very low top mass and light messengers its
effect can increase to $\sim \pm 2$. We also fixed $m_b(m_b) =
4.2\,\GeV$; changing $m_b(m_b)$ by $\pm 100\,\MeV$ typically changes
$\tb$ by $\mp 1$. The figures do show that the predicted $\tb$
increases slowly with the superpartner mass scale (since $m_A^2$
increases somewhat faster than $B\mu$ for various reasons). The $\tb$
prediction decreases as the top mass is increased. And as the
messenger scale is increased above $\Lambda$, at first $\tb$
increases somewhat {\it if} (contrary to our expectations) $\Mess3$
starts out degenerate with $\Lambda$ while $\Mess2 > \Lambda$, since
$M_3 \propto g(x_3)$ decreases rapidly as $x_3$ falls below unity, so
the dominant gluino contribution to $B^A$ is decreased and the
cancellation with $B^G$ which amplifies $\tb$ is strengthened; but
once the messenger scale is significantly above its minimum value,
raising it further will increase $L$, which increases $B^A$ more than
$B^G$, thereby reducing the cancellation and reducing $\tb$. The same
reasoning explains why $\tb$ is enhanced when, as expected, $\Mess3 >
\Mess2$ for light messengers, and why in that case $\tb$ decreases
monotonically as the messenger scale rises. When $\Mess{} \gg
\Lambda$, a small splitting between $\Mess3$ and $\Mess2$ makes
little difference. Most important, we reemphasize that the prediction
of $\tb$ depends completely on the assumption that $\Abart = \Abarb =
\Abartau = \Bbar = 0$.

For all the parameter values used in these figures, the $A$-term
contribution to $B$ was larger in magnitude than the gaugino
contribution, so $\tb$ (in our conventions) turned out positive.
However, the cancellation between $B^A$ and $B^G$ is a delicate one,
and by varying the parameters it is possible to find $\tb < 0 $
solutions (or even a positive and a negative solution for the same
parameters). The negative solutions tend to be smaller in magnitude,
$\sim -35$.  We find them only for extremal values of several
parameters, notably a light top quark and a very low messenger scale,
when the gaugino contribution has a chance of competing successfully
against the $A$ term contribution. Near these extremal parameter
values, the prediction of $\tb$ tends to be very large if positive, a
sign of the delicate cancellation, and is sensitive to contributions
to $B$ beyond those we have calculated (as well as to small
variations in the parameters of the model). We will not consider the
negative solutions further in this work, since we view them as
atypical and unlikely, and also because they lead to constructive
interference between the chargino and charged-Higgs amplitudes in the
process $b\to s\gamma$ (discussed below) and therefore require heavy
superpartner masses. We have, in any case, estimated the effects of
higher-order terms on the prediction of $\tb$. One source of
uncertainty is the uncalculated correction $c_\Delta$, but it tends
to cancel in the ratio $B\mu/m_A^2$ since $m_A^2$, $\mu$ and $B$ all
increase with $c_\Delta$ The result is an uncertainty in $\tb$ of
$\sim \pm 1$ for typical parameter values. More serious are other
possible higher-order terms we have omitted. To model their effects,
we vary the effective log $L_{A3Q}$ in the dominant gluino
contribution to $B^A$ by $\pm 1/L$. The predicted $\tb$ varies by
$\sim \pm 4$ when $\Mess{} = \Lambda$, by $\sim \pm (2-3)$ when
$\Mess{} = 2 \Lambda$, and by only $\sim \pm (1-2)$ when $\Mess{} =
10 \Lambda$. This is a significant uncertainty for low messenger
scales, but requires significantly more computation to eliminate, and
we will leave that for future work.

\bfig
\leavevmode
\epsfysize=15cm \epsfbox{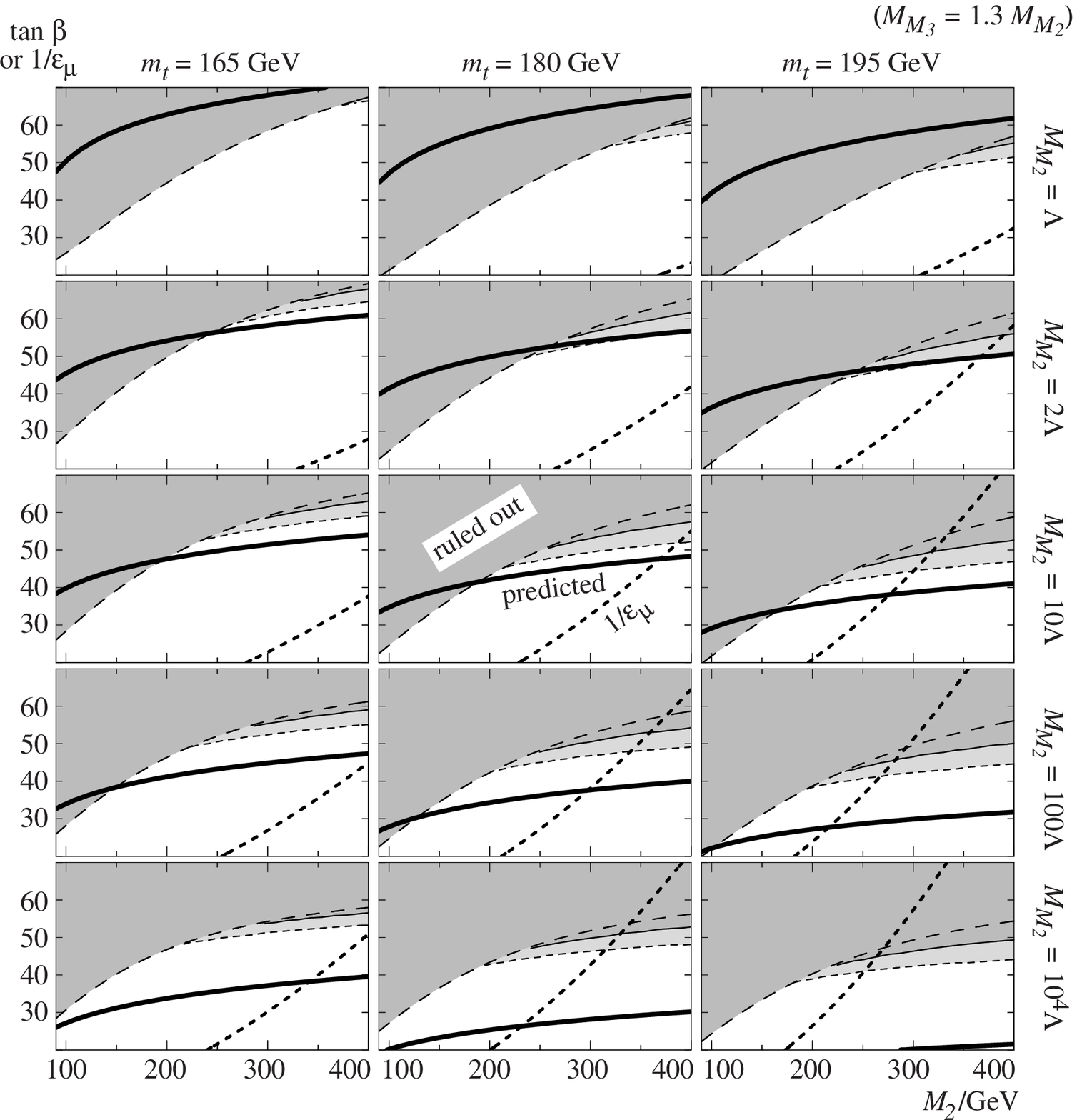}
\begin{quote}
{\small
\newfig \label{fig:bigone} The thick curves are our prediction for
$\tb$ as a function of the wino mass $M_2$ for various values of the
top pole mass $m_t$ and of the doublet messenger mass $\Mess2$,
fixing $\Mess3 = 1.3 \Mess2$ as roughly expected from RG evolution,
as well as $\alphas(m_Z) = 0.12$ and $m_b(m_b) = 4.2\,\GeV$. The
long-dashed curves are the upper bound $\tbz$ from the stau
mass-squared being positive and above $(\half m_Z)^2$; the solid
curves are the upper bound $\tbtbnd$ from the charge-conserving but
false vacuum being sufficiently long-lived; and the area they rule
out is shaded a dark gray. The more model-dependent bound $\tbc$ from
the mere existence of a deeper charge-breaking true vacuum appear as
short-dashed curves, and the region they ``rule out'' is shaded a
light gray. Finally, the fine-tuning measure $1/\epsmu$ is shown as a
heavy dotted curve.}
\end{quote}
\efig

\bfig
\leavevmode
\epsfysize=15cm \epsfbox{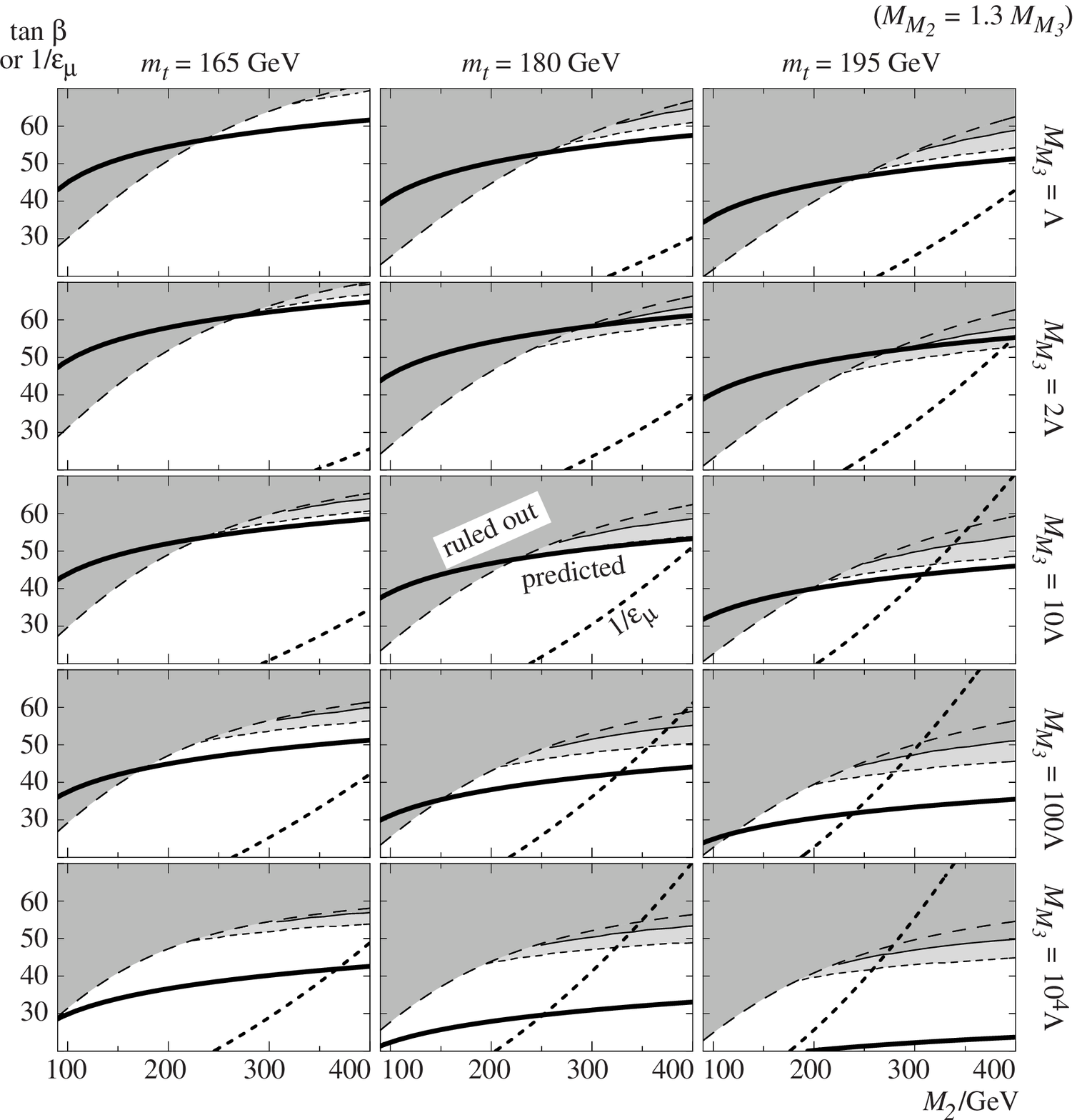}
\begin{quote}
{\small
\newfig \label{fig:bigtwo} Similar to Fig.~\ref{fig:bigone}, but
varying the triplet messenger mass $\Mess3$ while fixing the doublet
messenger mass to be slightly higher, $\Mess2 = 1.3 \Mess3$.}
\end{quote}
\efig

\section{Vacuum Stability I}

We turn now to less model-dependent features, and consider the
constraints imposed by requiring that the vacuum not break $\rm
SU(3)\times
U(1)_{em}$. We expect, as have others before us, the most severe
bounds to arise in the sleptonic sector, since the lowest-order
sleptonic
mass-squared parameters are small, in proportion to electroweak
couplings, and can be overwhelmed by destabilizing terms proportional
to the strong interactions. We will consider three such bounds: the
classical stability of the electroweak vacuum, or the improved
constraints from a lower bound on the lightest slepton mass; the
absence of a charge- or color-breaking minimum in the scalar
potential deeper than the ``ordinary'' electroweak-breaking minimum;
and a sufficiently small probability of quantum tunneling out of the
ordinary minimum and into a different charge- or color-breaking true
minimum, should one exist.

To derive these bounds, we start with the RG-improved tree-level
scalar potential and restrict our attention to the neutral component
of the up-type Higgs field $H_U$, the charged component of the
SU(2)-doublet stau field $\slep$ and the SU(2)-singlet stau field
$\stau$. We need only the up-type Higgs because the other Higgs has a
large and positive squared mass, and only the third generation of
sleptons because the destabilization arises from a large Yukawa
coupling. We find:
\bea
V &=& m_U^2 H_U^2 + \mslep^2 \slep^2 + \mstau^2 \stau^2 - 2 \ltau
\mu H_u \slep \stau + \ltau^2 \slep^2 \stau^2  \nonumber\\
& & \phantom{.} + {g_2^2\over8} \left(\slep^2+H_U^2\right)^2 +
{g_Y^2\over8} \left(\slep^2 - 2 \stau^2 - H_U^2\right)^2 +
\left({g_2^2 + g_Y^2\over8}\right) \delta_H H_U^4
\label{eq:V0}
\eea
which should be minimized when evaluated at a scale comparable to the
expectation values of the fields, namely, near the electroweak scale
or the superpartner mass scale. As before, we will actually evaluate
its mass parameters near $\msq$. The term $\propto \delta_H$ is the
leading correction from the full 1-loop effective potential.

The destabilizing term in Eq.~(\ref{eq:V0}) is the trilinear one, $-2
\ltau \mu H_u \slep \stau$, which can be large because $\mu \sim
\Mbar3$ and because a large $\tb$ requires a large $\ltau$ to
generate the known value of $m_\tau$:
\beq
\ltau = {m_\tau\over\vev\cos\beta} \simeq {\tan\beta\over100}\,.
\label{eq:ltau}
\eeq
We neglect the $\tb$-enhanced but weak finite threshold corrections
$\dmtau$. When $\tb$ is small, the trilinear term is negligible, the
potential has a minimum at $H_U = \vev$ and $\slep = \stau = 0$ and
no other minima, and the sleptons have masses $\roughly{>} \half
m_Z$. Let us define $H_U = \vev + H_U'$ and rewrite the potential in
terms of the shifted field $H_U'$; in addition to a trilinear term
$-2 \ltau \mu H_u' \slep \stau$ (and other cubic terms), the
sleptonic mass matrix (defined at the new origin $\langle H_U'\rangle
= \langle \slep\rangle = \langle \stau\rangle$) acquires a mixing
term $\mu \ltau \vev$. As $\tb$ is increased, the trilinear term
grows, while the lightest slepton mass-squared eigenvalue $\mlight^2$
decreases. If the trilinear term induces a new, discrete minimum in
$V$ while $\mlight^2$ is still positive, then both the electroweak
vacuum at the origin and the new vacuum are classically stable;
further increasing $\tb$ lowers the new vacuum and decreases
$\mlight^2$ until the latter goes negative and the origin is no
longer a minimum (but rather a saddle point). On the other hand, if
$\mlight^2$ goes negative while the trilinear term is too small, then
the vacuum at the origin is destabilized and a new vacuum moves
continuously away from the origin. These two situations are shown
schematically for a single field in Fig.~\ref{fig:bounds}. The actual
potential $V$ is of course a polynomial in three fields, but displays
qualitatively the same behavior. For small $M_2$ the origin is
classically destabilized first, leading to a single bound on $\tb$
(namely $\mlight^2 > 0$), while for larger $M_2$ the cubic term first
induces a new vacuum and only later is the origin destabilized, so
two possible bounds may be imposed. In fact the bound on $\mlight$
may be significantly (for light $M_2$) improved by imposing the
phenomenological LEP constraint $\mlight^2 > (\half m_Z)^2$; even
slightly stronger bounds are now available, but they will not change
our results appreciably. If $M_2$ is large and $\tb$ is large enough
so $V$ possesses a new minimum but small enough so the old minimum is
classically stable, we must still demand that the old minimum be
sufficiently stable to quantum tunnelling. The resulting bound is
discussed further below.  All these bounds are shown in
Fig.~\ref{fig:bounds} for a typical set of parameters $m_t =
180\,\GeV$, $\Mess2 = 10\Lambda$, $\Mess3 = 1.3 \Mess2$,
$\alphas(m_Z) = 0.12$ and $m_b(m_b) = 4.2\,\GeV$, and with $c_\Delta
= 0$. In the remaining we will use the $\mlight^2 > (\half m_Z)^2$
bound instead of $\mlight^2 > 0$.

\bfig
\leavevmode
\epsfysize=10cm \epsfbox{FigBOUNDS.eps}
\begin{quote}
{\small
\newfig \label{fig:bounds} Illustrating the various sleptonic vacuum
stability bounds for a particular choice of parameters. In the top
plot, the heavy long-dashed line corresponds to the stability of the
ordinary, charge-preserving vacuum: $\mlight^2 > 0$. This may be
improved to the phenomenological statement $\mlight^2 > (\half
m_Z)^2$ shown by the thin long-dashed curve. For sufficiently heavy
$M_2$ a new minimum develops away from the origin when $\tb$ exceeds
the short-dashed line. And when such a minimum exists, the lifetime
of the ordinary vacuum can be shorter than the observed age of the
universe when $\tb$ exceeds the solid line. The different vacuum
structures for light and heavy $M_2$ are shown schematicallyin the
two lower subplots.}
\end{quote}
\efig

\subsection{The Electroweak Vacuum}

We first find the upper bound $\tbz$ on $\tb$ above which $\mlight^2
< (\half m_Z)^2$. We need the sleptonic mass-squared parameters:
\beq
\mslep^2 \simeq {3\over2} \Mhat2^2 \left[
1 + {1\over5}\left({\alphbar1\over\alphbar2}\right)^2 +
\left\{{2 \alphbar2 g^2(x_2) - 2 \alphbar{\tau}\over2\pi}\right\} L
\right]
\label{eq:msl2}
\eeq
and
\beq
\mstau^2 \simeq {6\over5} \Mhat1^2 \left[
1 + \left\{{2 \alphbar1 g^2(x_1) - \alphbar{\tau} \left(3 + 5
\alphbar2^2/\alphbar1^2\right) \over 2\pi}\right\} L
\right]\,.
\label{eq:mstau2}
\eeq
The quartic terms in $V$ resulting from D-terms provide additional
mass-squared contributions to the stau mass matrix, which reads, in
the ordinary electroweak-breaking vacuum:
\beq
\staumat^2 =
\left(\begin{array}{cc}
\mslep^2 + (\half-\sw) m_Z^2 & \mu \ltau \vev \\
\mu \ltau \vev & \mstau^2 + \sw m_Z^2 \end{array}\right) \equiv
\left(\begin{array}{cc}
\mslepD^2 & \mu \ltau \vev \\
\mu \ltau \vev & \mstauD^2 \end{array}\right)\,.
\label{eq:massmat}
\eeq
Requiring the lightest eigenvalue to be sufficiently heavy then
amounts to
\beq
\mlight^2 = \half \left(\mslepD^2 + \mstauD^2 -
\sqrt{(\mslepD^2-\mstauD^2)^2 + 4
\left(\mu\ltau\vev\right)^2}\right) \roughly{>} (\half m_Z)^2\,.
\label{eq:staubound}
\eeq
By solving this constraint self-consistently for $\tb$ we obtain the
upper bound $\tbz$ shown in Figs.~\ref{fig:bigone} and
\ref{fig:bigtwo}. Note that this constraint weakens essentially
linearly with increasing superpartner mass, as the off-diagonal
entries in $\staumat^2$ are suppressed relative to the diagonal ones
by $\sim \vev/\Mhat{1,2}$.

\subsection{A New Charge-Breaking Vacuum}

But in fact the potential may become unacceptable even before this
point \cite{ref:us}, for $\tb < \tbz$, as long as $M_2$ is heavy
enough. As dicussed above, the scalar potential may develop a new
charge-breaking minimum for sufficiently large $\tb$ and $\mu$, which
quickly becomes deeper than the $\slep = \stau = 0$ vacuum. We denote
by $\tbc$ the critical value of $\tb$ at which this happens. When
$\tb$ just exceeds $\tbc$ the charge-breaking, true vacuum is just
deeper than the ordinary, false vacuum. This situation may still be
phenomenologically acceptable if the early universe is initially in
the ordinary vacuum, since the tunneling rate out of that false
vacuum is extremely small. However, some recent arguments suggest
that inflationary scenarios would inevitably leave the post-inflation
universe in the true vacuum, in which case we must ensure that the
ordinary vacuum is indeed the true one, that is, $\tb < \tbc$; the
issue seems contentious, model-dependent, and at present  unresolved
(see, e.g., Ref.~\cite{ref:toby} and references therein).  Such
cosmological arguments are beyond the scope of this work, so we will
calculate $\tbc$ and leave its significance to the reader's judgment.
In any case, as $\tb$ is increased beyond $\tbc$, the tunneling rate
from false to true vacuum increases until, at $\tb = \tbt$, the
lifetime of the false vacuum falls below the observed age of our
universe. Thus $\tbt$ is a firm upper bound on phenomenologically
acceptable values of $\tb$.

To see whether $V$ has a charge-breaking minimum, we insert the
expressions for $m_U^2$, $\mslep^2$, $\mstau^2$, $\mu$ and $\ltau$
given above into Eq.~(\ref{eq:V0}) and numerically search for a
minimum having $\slep \ne 0$ or $\stau \ne 0$; in fact, in the new
minimum $H_U$, $\slep$ and $\stau$ all receive comparable expectation
values. To be concrete, we consider a charge-breaking new minimum
harmless unless it's deeper than the ordinary one, so we impose:
\beq
\min V \ge \min_{\slep = \stau = 0} V\,.
\label{eq:vevbound}
\eeq
Since the new minimum grows much deeper than the ordinary minimum as
$\tb$ increases beyond $\tbc$, our result for $\tbc$ is very
insensitive to how deep we require the new minimum to be. The bound
$\tbc$ above which there is a dangerous new minimum is shown in
Figs.~\ref{fig:bigone} and \ref{fig:bigtwo} as a function of $M_2$.
We only display $\tbc$ when it is below $\tbz$, that is, when the
former is a stronger constraint. But we continue to display $\tbz$
even when it is weaker than $\tbc$ because the existence of a deeper
minimum {\it may} not be unacceptable, depending on whether the
cooling, expanding universe can be trapped in the false vacuum and on
how long it would take to tunnel into the true vacuum.

\subsection{Tunneling into the New Vacuum}

We will not address the cosmological issues which determine into
which vacuum the universe cools---rather, we treat $\tbc$ as the
value above which $\tb$ {\it may} be unacceptable. It is {\it
definitely} unacceptable, though, if the universe would have tunneled
out of the ordinary, false vacuum in less than $\sim10^{10}$ years
even if it did start out there. To compute the lifetime of the false
vacuum, that is, the time interval over which the tunneling
probability becomes essentially unity, we use the semiclassical
bounce approximation  \cite{ref:coleman}. The probability, per unit
volume and per unit time, of forming a bubble of true vacuum in a
false vacuum universe is
\beq
\Gamma/V = A e^{-\SEb}
\label{eq:Gamma}
\eeq
where $A$ is a product of masses and couplings in $V$ and some
additional numerical factors; we expect it to be very roughly the
fourth power of the electroweak scale or the superpartner masses. The
tunneling rate is much more sensitive to the exponent $\SEb$, which
is the Euclidean action of the ``bounce'', a particular
O(4)-invariant solution to the classical Euclidean equations of
motion for the fields in $V$. Requiring the probability of tunneling
over the present age of the universe, and within the observable
volume of the universe, to be much less than unity implies $\SEb
\roughly{<} 400$. For a given $\tb$ we need to construct the bounce
solution, namely find the configuration of the three fields $H_U$,
$\slep$ and $\stau$ which extremizes $S_E$ and approaches the false
vacuum configuration at infinite Euclidean time; the maximal value of
$\tb$ for which the extremal value of $S_E$ does not exceed 400 would
be $\tbt$, and higher values of $\tb$ would definitely be ruled out.

When the bounce configuration involves several fields and a
complicated potential, the only available analytical solution is in
the thin-wall approximation. This is valid when the true vacuum is
only slightly deeper than the false one, namely when $\tb \sim \tbc$,
and the lifetime of such a marginally-unstable universe is inevitably
extremely long. Therefore to find the bounce solution for which the
universe lives only $10^{10}$ years requires numerical extremization
of the action. Rather than embarking on a numerical extremization on
the lattice (using for example the methods of
Refs.~\cite{ref:lattice}), we will only compute an upper bound on the
bounce action, and hence an upper bound $\tbtbnd$ on the maximal
allowed value of $\tb$. To find an upper bound on $\SEb$, we choose a
ray in the three-dimensional space of our fields $H_U$, $\slep$ and
$\stau$, originating in the false vacuum and extending in some
direction along which $V$ is eventually deeper than the false
minimum. We calculate the one-dimensional bounce action in this
direction, and then vary the direction to minimize $S_E$ (as was also
done for a different potential by Dasgupta, Dobrescu and Randall
\cite{ref:ddr}). The minimum value of this straight-line action is an
upper bound on $\SEb$ \cite{ref:sidney}. For the potential and
parameter values we're studying, it turns out that the straight-line
action connecting the false to the true vacuum yields essentially the
lowest tunnelling action, so scanning over the various directions
yields little improvement, lowering $\tbt$ by $\sim 2\%$. A
semi-analytic calculation of this bounce action for any quartic
potential, a discussion of why it bounds the bounce action from
above, and further details about our particular potential are
presented in Ref.~\cite{ref:qb}. Using those results, we obtain the
upper bound $\tbtbnd$ shown in Figs.~\ref{fig:bigone} and
\ref{fig:bigtwo} as a function of $M_2$. Just as for the bound $\tbc$
from the absence of a new minimum, we only display the tunneling
bound $\tbtbnd$ when it is stronger than the $\tbz$ constraint;
unlike the $\tbc$ bound, however, the tunneling bound is an absolute
one, which obviates the $\tbz$ bound when $\tbtbnd < \tbz$. (We still
display $\tbz$ even when it's weaker to show the improvement.) We
expect that the tunneling bound could be somewhat strengthened by
improving the bounce calculation beyond the straight-line
approximation, in which case $\tbtbnd$ would be lowered closer to the
true tunneling bound $\tbt$ and hence closer to $\tbc$.

The three sleptonic bounds derived above --- from requiring the
ordinary $\slep = \stau = 0$ vacuum to be a local minimum with a
sufficiently heavy stau ($\sim\tbz$), or to be a global minimum
($\tbc$), or to be a local but long-lived minimum ($\tbt$ or
$\tbtbnd$) --- carry an implicit uncertainty because of the parameter
values used in evaluating $V$. An important source is the
uncalculated higher-order correction $c_\Delta$ in $\Dt^2$ and
therefore in $\mu$. Varying $c_\Delta$ between $\pm 1/L_\Delta$
changes the bounds on $\tb$ by $\sim \mp 5$ when $L_\Delta \simeq
\Lmin = \Lambda$, and by $\sim \mp (1-2)$ when $L_\Delta \simeq 100
\Lambda$. This is a serious uncertainty when the messengers are
light, and impedes the comparison of the bounds with the predicted
values of $\tb$. The situation is exacerbated by the uncertainty
remaining in those $\tb$ predictions, discussed above. Since the
predictions and the bounds are rather close over a wide range of the
model's parameters, it is difficult to draw precise quantitative
conclusions regarding which parameter ranges are excluded. We will
address what can be learned from this comparison in the concluding
paragraphs.

\section{Vacuum Stability II: Bounding the Messenger Scale}

Can the messenger scale be increased arbitrarily high? We think not,
based on cosmological arguments limiting the mass of the gravitino,
and also based on our analysis of another charge-breaking bound,
first noticed by Komatsu \cite{ref:komatsu}.

If the gravitino is extremely light and was thermalized in the early
universe, its relic density {\it could} exceed the critical density
(and hence unacceptably shorten the universe's lifetime) unless its
mass obeys $m_{3/2} \roughly{<} 1\,\keV$ \cite{ref:papri}. An upper
bound on $m_{3/2}$ yields an upper bound on the messenger scale
\cite{ref:ambro,ref:damien}. Specifically, $m_{3/2} = \langle F_{\rm
max}\rangle/\Mpl$ where $\langle F_{\rm max}\rangle \geq \vevFS$ is
the order parameter of supersymmetry breaking, and $\vevFS$ is the
$F$ term of the singlet which gives mass to the messengers. Therefore
the bound on the messenger mass is $\Mess{i}/\Lambda < \lambda_i
[\Mpl (1\,\keV)/\Lambda^2] \simeq \lambda_i 10^4 (100\,\GeV/M_2)^2$.
Depending on the messenger Yukawa couplings $\lambda_i$ and the
superpartner mass scale, the upper bound on $\Mess{i}$ can range from
$\sim 10^4 \Lambda$ to much smaller values. However, as studied in
Ref.~\cite{ref:mormur}, a late period of inflation with a
sufficiently low reheating temperature could dilute the gravitinos
and relax the bound by many orders of magnitude.

An upper bound on the messenger scale also follows from requiring the
absence of charge-breaking minima along the flat directions of the
MSSM scalar potential. There are only two directions in field space
(restricted to the third generation and Higgs sectors) for which the
quartic terms in the scalar potential vanish: (i) $\langle H_U
\rangle = \langle H_U \rangle = \phi_1$ with all other fields
vanishing, and (ii) $\langle H_U \rangle \equiv \phi_2$, $\langle
\widetilde L\rangle = \sqrt{\phi_2^2 + \phi_2 \mu/\lb}$, and $\langle
\widetilde Q\rangle = \langle \widetilde b\rangle =
\sqrt{\phi_2\mu/\lb}$, where color and isospin orientations are
determined by requiring vanishing D-terms along this direction. Along
the first direction the potential is purely quadratic: $V(\phi_1) =
m_1^2 |\phi_1|^2$; the resulting stability constraint is the
well-known one, $m_A^2 > \mu B$, which in our case is always
satisfied. The second direction, first noted by Komatsu
\cite{ref:komatsu}, was discussed in some detail in
Ref.~\cite{ref:us} in the context of large $\tb$; our analysis here
will be essentially equivalent. The potential along $\phi_2$ is
\bea
V(\phi_2) &=& \left(m_{H_u}^2 + \mslep^2\right) |\phi_2|^2 +
\left(\mslep^2 + \msq^2 + \msb^2\right)
\left|\mu\phi_2\over\lb\right|
\nonumber\\
&\equiv& m_2^2 |\phi_2|^2 + m_3^2 \left|\mu\phi_2\over\lb\right|\,.
\label{eq:Vphitwo}
\eea
To check whether this potential has a minimum at some field value
$\phi_2$, we should evaluate its parameters at a scale $\sim \phi_2$.
For field values below $\Lambda_{\rm HIGH} \sim 10^7 - 10^9 \,\GeV$,
any higher-dimensional operators, and in particular a neutrino-mass
operator, would be ineffective in stabilizing this potential.
Therefore we must ensure $V$ does not develop a minimum at any scale
below $\Lambda_{\rm HIGH}$. (Thus $\Lambda_{\rm HIGH}$ is an upper
bound on the expectation value of $\phi_2$, which would become
relevant should the messenger scale exceed $\Lambda_{\rm HIGH}$, but
we will argue below that that is unlikely.) The potential could
develop a charge-breaking minimum if $m_2^2$ becomes negative as a
result of RG evolution below the messenger scale. This is a real
possibility, since the up-type Higgs SUSY-breaking mass $m_{H_u}^2$
($ = m_U^2 - \mu^2 \simeq -\half m_Z^2 - \mu^2$) becomes negative and
of order $M_3^2$ at low scales, whereas $\mslep^2$ is only of order
$M_2^2$. But if $m_2^2$ is driven negative only at very low scales,
the linear term in $V(\phi_2)$ can stabilize the potential and
prevent the appearance of a new, and possibly true, vacuum. To study
whether a new vacuum develops at  some field value $\phi_2$, we
evolve $V(\phi_2)$ starting from the messenger scale down to the
scale $\phi_2$. To a first and sufficient approximation, we only
keep\footnote{The potential along the flat direction $\phi_2$ is
determined in leading-$\log(\phi_2)$ approximation by the tree
potential evaluated at a field-dependent scale $\phi_2$
\cite{ref:us}.  For our purposes, this dependence on the log just
corresponds to the first step in the solution to the 1-loop RG
equation for $m_2^2$. The result has the following effective-theory
interpretation. Along $\phi_2$ a set of states gets masses $\sim
\phi_2$ while other states are much lighter ($\sim \msusy$ or
$\sim\sqrt{\phi_2\msusy}$). The contribution of the heavy states to
the 1-loop effective potential is the field-dependent correction to
$m_2^2$, namely the $\Delta m_2^2 |\phi_2|^2 \log \phi_2$ term below
(after taking into account the supersymmetric cancellation of the
leading $\phi_2^4$ terms). The lighter states contribute at order
$\msusy^4$ or $\msusy^3\phi_2$ to the potential, and we will drop
them. These terms would affect the renormalization of $m_3$, for
which we are content with the lowest result, as long as it is not
zero.} the leading correction to $m_2^2$ and ignore the small
corrections to $m_3^2$, $\mu$ and $\lb$. We minimize
\beq
V(\phi_2) \simeq \mbar2^2 |\phi_2|^2 + \Delta m_2^2 |\phi_2|^2 \ln
{\phi_2\over\Mess{}} + \mbar3^2 \left|\mu\phi_2\over\lb\right|
\label{eq:Vphitwoa}
\eeq
where $\Delta m_2^2 \simeq (2/\pi^2) \lbart^2 \Mhat3^2$, and either
require that the resulting equation has no solution, so no minimum
develops away from the origin, or that the new minimum is not deeper
than $V(\phi_2 = 0) = 0$. We find that as long as the messenger scale
is below a critical value
\beq
\Mcrit = \left\{\half e^{3/2}\,{\rm or}\,e\right\}
\left({\mbar3^2\mu\over\Delta m_2^2\lb}\right)
e^{\mbar2^2/\Delta m_2^2}
\label{eq:Mcrita}
\eeq
then $V(\phi_2)$ has no minimum (when the $\half e^{3/2}$ prefactor
is used) or at least has no negative minimum (when the $e$ prefactor
is used). Notice that the critical messenger mass is quite sensitive
to the exact parameters used in the potential, a typical feature for
dimensional transmutation. Our approximation will suffice for a {\it
rough} upper bound on the messenger mass. Evaluating the various
quantities in Eq.~(\ref{eq:Mcrita}) to leading order, we find
\beq
\Mcrit \simeq  \mu \left(\left\{\half e^{3/2}\,{\rm or}\,e\right\} 8
 \pi^2 \over 3 \lb \lt^2\right)^{1/(1-q)} e^{p/(1-q)}
\label{eq:Mcritb}
\eeq
where
\beq
p \equiv {3\pi^2\over 2\lt^2} \cdot
{\alpha_2^2(\mu)\over\alpha_3^2(\mu)}
\label{eq:peq}
\eeq
and
\beq
q \equiv {3\pi\over 4\lt^2} \cdot
{\alpha_2^2(\mu)\over\alpha_3^2(\mu)} \cdot
\left(\frac{34}{2} \alpha_3 - 6 \alpha_t\right)\,.
\label{eq:qeq}
\eeq
When $M_2 = 100\,\GeV$, the critical messenger scale ranges from
$\Mcrit/\Lambda \sim 300$ for $m_t = 165\,\GeV$ to $\Mcrit/\Lambda
\sim 10$ for $m_t = 195\,\GeV$; when $M_2 = 400\,\GeV$, it ranges
from $\Mcrit/\Lambda \sim 6\times 10^4$ to $\Mcrit/\Lambda \sim 50$
for the same top mass variation. This rough estimate of the upper
bound on the messenger scale can therefore be stronger than,
comparable to, or weaker than the cosmological bound from the
gravitino mass.

How serious is this $\Mcrit$ upper bound on the messenger scale? As
was the case for our previous vacuum-stability bound, the answer
depends on cosmological arguments beyond the scope of our work. If
$\Mess{} > \Mcrit$, a charge-breaking true vacuum exists far along
the flat direction $\phi_2$. However, as was shown approximately in a
similar context in Ref.~\cite{ref:strumia} and as we calculate more
accurately in Ref.~\cite{ref:qb}, the tunneling rate into this
unacceptable vacuum is extremely small, that is, the lifetime of our
familiar but false vacuum is much longer than the present age of the
universe. Thus $\Mcrit$ constitutes an upper bound on the messenger
mass only if the true vacuum could be reached directly as the early
universe cooled, rather than by tunneling in the cold universe.
Recent studies \cite{ref:toby} suggest that at finite temperatures
the true vacuum along the flat direction is indeed favored, but at
present this question remains far from settled.

\section{$b\to s\gamma$}

An interesting phenomenological consequence of a large $\tb$, as
shown in Ref.~\cite{ref:hrs} (see also Refs.~\cite{ref:bsg}), is
closely related to the large $\dmb$ radiative corrections to the
bottom quark mass. Diagrams very similar to those which yield $\dmb$
contribute to the radiative decay $b\to s\gamma$ in proportion to
$\tb$, when a photon line is attached wherever possible. The three
diagrams of interest involve the exchange of a higgsino, a mixed
wino-higgsino, or a gluino. The first is induced directly by the
flavor mixing $|V_{ts}| \simeq |V_{cb}|$ in the quark sector, while
the latter two require a squark mixing angle $\Vtt$ between the
SU(2)-doublet stop and scharm fields. We work in a supersymmetric
flavor basis where the down-type Yukawa matrix is diagonal, so that
all CKM mixings are contained in the up-type Yukawa matrix. If
$\mtt^2$ is the off-diagonal mixing element in the SU(2)-doublet stop
and scharm mass-squared matrix, while $\msq^2$ is the average of the
almost-degenerate diagonal masses, we define $\Vtt \equiv
\mtt^2/\msq^2$ (denoted by $V_{23}$ in Ref.~\cite{ref:us}). Since
$A$, $B$ and $\mtt^2$ are radiatively induced in the model we are
considering, we can calculate the sign and magnitude of all three
amplitudes. (This is qualitatively different from the
gravitationally-mediated SUSY-breaking models considered in
Ref.~\cite{ref:us}, in which $B$ and $\mtt^2$ were free parameters,
so the sign of the somewhat dominant higgsino-exchange diagram was
inferred from the phenomenological constraint $\dmb < 0$, while the
other two diagrams were largely unknown.)

When the large QCD corrections are taken into account, the branching
ratio for $b\to s\gamma$ is proportional to:
\bea
{\rm BR}(b\to s\gamma) &\propto&
\left|\eta^{16/23} C_7^{(0)} +
\frac83\left(\eta^{14/23}-\eta^{16/23}\right) C_8^{(0)} + C_2^{(0)}
\sum h_i \eta^a_i \right|^2
\label{eq:eBRbsg}\\
&\simeq& \left|0.70 C_7^{(0)} - 0.166\right|^2
\label{eq:aBRbsg}
\eea
where in the first line we have used the leading-log expressions
summarized in Ref.~\cite{ref:Bur} and in the second line we have
evaluated them using  $\eta = \alpha_s(m_W)/\alpha_s(\mu \sim m_b)
\simeq 0.59$ (when $\alpha_s(m_z) = 0.12$). The next-to-leading-order
calculation has now been essentially completed \cite{ref:misiak},
with the result that there are no large corrections to the
leading-order calculation, and hence only its uncertainties are
reduced. We will continue to parametrize deviations from the standard
model predictions as deviations in $C_7^{(0)}$, since they will turn
out to be potentially very large and so an estimate is all we need.
The amplitude $C_7^{(0)}$ is given in the standard model by the
W-exchange amplitude $\A_{\rm SM}$, while in the MSSM there is an
additional contribution $\A_{\rm H^-}$ due to charged-Higgs exchange
and the three amplitudes $\A_{\rm \widetilde h}$, $\A_{\rm \widetilde
W \widetilde h}$ and $\A_{\rm \widetilde g}$ discussed above. They
are well-approximated by
\bea
\A_{\rm SM} &=& \frac32 {m_t^2\over m_W^2}
f_\gamma^{(1)}(m_t^2/m_W^2)
\label{eq:ASM}\\
\A_{\rm H^-} &=& \frac12 \left({1\over 1+\dmb}\right) {m_t^2\over
m_{H^-}^2} f_\gamma^{(1)}(m_t^2/m_{H^-}^2) \label{eq:A2H}\\
\A_{\rm \widetilde h} &=& \frac12 \left({\tb\over 1+\dmb}\right)
{m_t^2 A_t \mu\over \msq^4} f_{\rm ch}(\mu^2/\msq^2)
\label{eq:AHH}\\
\A_{\rm \widetilde W \widetilde h} &=& {m_W^2\over\msq^2} {\Vtt\over
V_{ts}} \left({\tb\over 1+\dmb}\right) {M_2 \mu\over M_2^2 - \mu^2}
\left[f_{\rm ch}(M_2^2/\msq^2)  - f_{\rm ch}(\mu^2/\msq^2)\right]
\label{eq:AWH}\\
\A_{\rm \widetilde g} &=& \frac89{\alpha_s\over\alpha_W}
{m_W^2\over\msq^2} {\Vtt\over V_{ts}} \left({\tb\over 1+\dmb}\right)
{M_3 \mu\over \msq^4} f_{\rm gl}(M_3^2/\msq^2)
\label{eq:AGL}
\eea
where
\bea
f_\gamma^{(1)}(x) &=& {7 - 5 x - 8 x^2 \over 36 (x - 1)^3} + {x (3 x
- 2) \over 6 (x-1)^4}\ln x
\label{eq:fg1}\\
f_\gamma^{(2)}(x) &=& {3 - 5 x \over 6 (x - 1)^2} + {3 x - 2 \over 3
(x-1)^3}\ln x
\label{eq:fg2}\\
f_{\rm ch}(x) &=& {13 - 7 x \over 6 (x - 1)^3} + {3 + 2 x (1 - x)
\over 3 (x-1)^4}\ln x
\label{eq:fch}\\
f_{\rm gl}(x) &=& {1 + 10 x + x^2 \over 2 (x - 1)^4} - {3 x (1 + x)
\over (x-1)^5}\ln x\,.
\label{eq:fgl}
\eea
To evaluate these we need, in addition to $\mu$, $\tb$, and $A_t$
calculated above, the approximate diagonal squark mass:
\beq
\msq^2 \simeq \frac83 \Mhat3^2  + \frac32 \Mhat2^2
\label{eq:msq2}
\eeq
and the scharm-stop off-diagonal mass-squared calculated to lowest
order using the RG evolution:
\beq
\Vtt \equiv {\mtt^2\over\msq^2} \simeq - {\lt^2\over4\pi^2} V_{ts}
L\,.
\label{eq:Vtt}
\eeq
In this last expression we have ignored effects from the charm mass
and renormalizations of the diagonal elements of the squark mass
matrix. Also, to lowest order there is no flavor mixing in the
SU(2)-singlet squark masses. The standard-model $b\to s\gamma$
amplitude has the same sign as the charged-Higgs-exchange amplitude.
When $B$ is dominated by the $A$ term contribution (as is almost
inevitable) and is therefore negative, we find that the higgsino and
mixed wino-higgsino diagrams have the same sign, which is {\it
opposite} to the standard-model and charged-Higgs amplitudes; the
much smaller gluino diagram has the same sign as the standard-model
amplitude.

To convert these amplitudes into an experimental prediction, we
define as in Ref.~\cite{ref:dtw} the amplitude deviation of the MSSM
relative to the standard model:
\beq
R_7 \equiv {C_7^{(0),{\rm MSSM}}\over C_7^{(0),{\rm SM}}} - 1
= {\A_{\rm H^-}\over \A_{\rm SM}} + {\A_{\rm \widetilde h} + \A_{\rm
\widetilde W \widetilde h} + \A_{\rm \widetilde g}\over \A_{\rm
SM}}\,.
\label{eq:R7}
\eeq
We find that the answer is sensitive mainly to two parameters: the
superpartner mass scale parametrized by $\Mhat2$, and the messenger
scale $\Mess{}$. In Fig.~\ref{fig:bsg} we show the two terms which
constitute $R_7$, with the charged-Higgs relative contribution as a
dotted line and the superpartner relative contributions as a dashed
line, all for a messenger scale $\Mess{} = 2 \Lambda$. While $\A_{\rm
H^-}$ adds to the standard contribution and has been used in the past
to bound the charged-Higgs mass, the superpartner diagrams as a whole
interfere destructively with the standard model and charged-Higgs
amplitudes and can even overwhelm them for sufficiently light
superpartners. The summed deviation $R_7$ is plotted as the solid
thick black line for $\Mess{} = 2 \Lambda$, and as the two thin black
lines for $\Mess{} = 10 \Lambda$ (upper) and for $\Mess{} = 100
\Lambda$ (lower). We see that $R_7$ is large and negative when the
superpartners are light, and even more so when the messengers are
heavy (since a higher messenger scale means larger $A$ terms are
generated and a larger $\mu$ is needed, and these enhance the
higgsino diagrams but suppress the charged-Higgs one). For
sufficiently heavy superpartners the charged-Higgs amplitude
dominates [since it is enhanced by $\log(m_A/m_t)$] and causes $R_7$
to tend towards 0 from above. Of course, these results should be
combined with the bounds derived above from vacuum-stability
conditions: light superpartners, with $M_2 < 200\,\GeV$, are only
allowed when the messengers are heavy, and preferably when the top
quark is heavy and the triplet messengers are heavier than the
doublet ones.

\bfig
\leavevmode
\epsfysize=7cm \epsfbox{FigBSG.eps}
\begin{quote}
{\small
\newfig \label{fig:bsg} The $b\to s\gamma$ amplitude deviation $R_7$
from the standard-model prediction (left), and the branching ratio
${\rm BR}(b\to s\gamma)$ multiplied by $10^4$ (right). The
charged-Higgs contribution to $R_7$ and to the branching ratio are
shown as the upper heavy dotted curve, while the chargino and gluino
contributions to $R_7$ are shown as the lower dashed curve, both for
$\Mess2 = 2 \Lambda$ (and $\Mess3 = 1.3 \Mess2$). The sum $R_7$ for
the entire MSSM and the resultant branching ratio are shown as the
heavy solid curves. The lighter curves correspond to increasing the
messenger scale to $\Mess2 = 10 \Lambda$ or $\Mess2 = 100 \Lambda$.
The standard model (SM) predictions are sown as thin straight lines.
The present CLEO bounds on $R_7$ are inferred using the
next-to-leading-order standard-model calculation, as described in the
text; the areas excluded at the 95\% CL are shown in gray, as is the
central value of the experimental result.}
\end{quote}
\efig

Our aproximation is sufficient to calculate the amplitudes to within,
say, $\sim 20\%$. To improve on this would require various new
calculations. For example, the dominant chargino contribution arises
from the generation of the $A_t$ term, but to calculate the
subleading effects would require calculating the full 2-loop diagram
in which the photon is attached not only to the squarks and chargino
but also to the top quark in the loop which generates $A_t$. This and
other theoretical uncertainties would need to be removed if finer
comparisons with experiment are to be made, once the experimental
uncertainty is substantially decreased.

Our results may be compared with previous calculations found in
Refs.~\cite{ref:dtw} and \cite{ref:desh}. Neither of these use the
$\Bbar = 0$ to calculate $\tb$ or the sign of the MSSM diagrams, so
they consider both constructive and destructive interference. Since
in Ref.~\cite{ref:dtw} $\tb$ is at most 20, it is difficult to
compare their results with ours; they find that the chargino
contributions are much smaller than the charged-Higgs one. The
authors of Ref.~\cite{ref:desh} consider the value $\tb = 42$, and we
find rough qualitative agreement with their results, although they
only consider heavy superpartners and do not present enough of their
assumptions and intermediate results to allow a simple comparison
(though they do not seem to use $\dmb$). They do, however, observe
(for very heavy messengers) the large destructive interference we
have found. We find this destructive interference for all messenger
scales, and as a result all messenger scales are agreement with
current experimental bounds on ${\rm BR}(b\to s\gamma)$, as we now
show.

To compare our results with experimental findings, we will use the
standard-model next-to-leading calculation of Ref.~\cite{ref:misiak}
and Eq.~(\ref{eq:aBRbsg}) to obtain:
\beq
{\rm BR}^{\rm MSSM}(b\to s\gamma) = (3.3 \pm 0.3)\times 10^{-4}
\left|1 + 0.46 R_7\right|^2\,.
\label{pBRbsg}
\eeq
(Using the older, less precise calculations of earlier references,
e.g. Ref.~\cite{ref:greub}, would not make much difference here,
since we are only aiming for an estimate of the effect of large $\tb$
in the MGM.)  This prediction should be compared with the present
experimental bounds of Ref.~\cite{ref:CLEO}: $1.0\times 10^{-4} <
{\rm BR}(b\to s\gamma) < 4.2\times 10^{-4}$ at 95\% CL. The resulting
allowed range of the deviation in amplitudes, $-1.0 < R_7 < 0.4$, is
shown as the white region in Fig.~\ref{fig:bsg}. We learn that at
present the GMSB (and indeed also the standard model) predictions are
consistent with experimental bounds on $b\to s\gamma$ for all
superpartner masses. However, if the branching ratio is measured much
more precisely, as expected, then strong limits may be placed on the
superpartner masses and perhaps also on the messenger scale. It is
intriguing that the current CLEO measurement \cite{ref:CLEO} ${\rm
BR}(b\to s\gamma) = (2.32 \pm 0.57 \pm 0.35) \times 10^{-4}$ is one
to two standard deviations {\it below} the standard model prediction.
Its central value corresponds to $R_7 \sim -0.35$, which is the
prediction of the $\Bbar = 0$ scenario when $M_2 \sim 150\,\GeV$. (Of
course, more accurate theoretical calculations would also be needed
to sharpen the prediction.)  If the branching ratio remains
significantly lower than the standard model prediction, then the
small $\tb$ GMSB scenario would be disfavored, while the $\Bbar = 0$
large $\tb$ scenario could prove very successful --- once a low
chargino mass and a high messenger scale are confirmed.

\section{Conclusions}

We have analyzed some aspects of a minimal gauge-mediated (MGM)
SUSY-breaking scenario characterized by a large hierarchy of up-
versus down-type Higgs VEVs: $v_U/v_D = \tb \gg 1$. Such a hierarchy
will arise naturally in this model provided the tree-level $B$
parameter (effectively) vanishes at the messenger scale. Therefore,
if some messenger-scale mechanism leads to $\Bbar = 0$, the
top-bottom hierarchy can {\it automatically} follow even if $\lt \sim
\lb$, in sharp contrast to the gravitationally-mediated scenario in
which large $\tb$ must be imposed through an artificial cancellation
between the high-scale value of $B$ and its large radiative
corrections. We have analytically calculated the radiative
corrections to $B$ in the MGM using an explicit perturbative
expansion in the gauge and Yukawa couplings, which furnishes a
prediction for $B$ once $\Bbar = 0$ is assumed, and allows estimation
of the uncertainties in this prediction. Such a careful expansion is
necessary because these radiative contributions not only arise at a
high loop order, but furthermore accidentally cancel at leading order
when the messenger scale is low ($\Mess{} \simeq \Lambda$). As a
result $\tb$ is predicted to be between roughly 40 and 60 for light
messengers, but decreases logarithmically with the messenger mass. We
have shown the detailed predictions for $\tb$ as a function of the
top quark mass (with which it decreases) and of the superpartner mass
scale (with which it increases). The remaining uncertainties in $\tb$
are between $\pm 4$ for light messengers to less than $\pm 1$ for
heavy messengers. As a by-product of this calculation, we obtained an
analytic approximation for the $\mu$ parameter value needed for
proper electroweak symmetry-breaking and also for the $A$ terms, and
we presented a measure of how precisely $\mu$ needs to be tuned to
this value as a function of the superpartner mass scale.

We then contrasted this prediction with some constraints arising from
the possibility of destabilizing the slepton-Higgs scalar potential
when the Yukawa coupling $\ltau$ is large. We derived two strict
bounds --- the ordinary charge-preserving vacuum must be at least a
local minimum with a sufficiently heavy stau, and if it is not a
global minimum then it must be at least as long-lived as the observed
age of the universe --- and a more cosmologically model-dependent
bound based simply on the existence of a deep charge-breaking true
vacuum. The first, well-known constraint (which we slightly improved
by demanding that the lightest slepton mass eigenvalue exceed
experimental bounds) dominates when the superpartners are light, and
indeed rules out significant portions of the parameter space. The
latter two constraints eventually dominate when the superpartners are
heavy, but are not as effective in eliminating portions of parameter
space. Unfortunately, they too are plagued by residual uncertainties
comparable to those in the prediction of $\tb$. Thus it is difficult
to precisely delineate which values of $m_t$, $\Mess{i}$ and $M_2$
(and thereby the remaining superspectrum) are allowed and which are
forbidden. To do so would require a substantial calculational effort
to obtain the next order in the perturbative expansion.
Qualitatively, though, it is clear that heavier messengers and heavy
superpartners allow the predicted $\tb$ to more easily satisfy the
constraints. Heavier messengers also produce a smaller $\tb$, so if
large $\tb$ is desired (for example to allow a degree of
third-generation Yukawa unification) then heavier superpartners are
the preferred option, in which case the vacuum-stability bounds are
the dominant restriction.

But raising the superpartner and messenger mass scales exacts a
price. The former would make it hard to understand why the
electroweak scale is so light: the amount of fine-tuning of $\mu$,
quantified by the parameter $1/\epsmu$ plotted in
Figs.~\ref{fig:bigone} and \ref{fig:bigtwo}, increases sharply when
$M_2$ rises, as well as when $m_t$ and $\Mess{}$ rise. And raising
the messenger scale can lead to cosmological difficulties due to
gravitino overabundance, or, as we showed in this work, due to vacuum
instability in the squark-Higgs scalar potential when $\lb$ is large.
While such cosmological arguments are fraught with unsettled issues,
they suggest that the messenger scale should not be too many orders
of magnitude above its minimum value $\sim \Lambda$.

It thus emerges that the large $\tb$ scenario can at least be roughly
bounded in all its parameters. It is even more predictive than the
generic MGM model, and offers several phenomenological dividends: a
natural solution of the SUSY CP problem, and progress towards an
explanation of the top-bottom quark mass hierarchy. We believe it is
worthy of further phenomenological analysis. One such direction we
have pursued in this work is radiative bottom-quark decay: $b\to
s\gamma$. The branching ratio we predict may be markedly lower than
both the standard model and low-$\tb$ MSSM predictions if the
superpartners are relatively light, and though no definitive
statements can be made until experimental uncertainties are reduced,
the large $\tb$ prediction is indeed in better agreement with the
current measurement!

Several possible directions for future study emerge from the present
work. First, the $\Bbar = 0$ predictions of $\tb$ are currently
plagued by significant uncertainties arising from the next order in
our perturbative expansion, while sharpening the more
model-independent bounds on $\tb$ requires calculating $c_\Delta$ to
pin down $\mu$. The calculation of this threshold is much more
involved than the $1+T(x)$ threshold we have calculated in this
paper, since the latter corresponds to the finite part of a single
2-loop diagram (with the topology of Fig. 1), while the former is the
finite part of a whole class of 3-loop diagrams. Both would be
important to another future study: whether the $b$, $\tau$ {\it and}
$t$ Yukawa couplings could be unified in a grand-unified theory such
as SO(10). A detailed study of complete third-generation Yukawa
unification in GMSB, parallel to the one of
Refs.~\cite{ref:hrs,ref:us} for the gravitationally-mediated
scenario, remains to be done and would require more refined threshold
calculations, but preliminary indications \cite{ref:GMSBYU} seem
promising. Another interesting direction would be to apply the
techniques presented in this work to a broader class of GMSB models,
in which the $\mu$ problem may be solved with possible modifications
of the Higgs masses, or in which the messenger sector is enlarged. As
long as such models feature a small $\Bbar$, say ${\cal O}(\alpha)$
where $\alpha$ is some small coupling, we expect that the ideas and
methodology of the present work, if not the quantitative results,
could still be applicable.

\section*{Acknowledgements}

We are grateful for stimulating conversations with Alessandro
Ambrosanio, Andrea Brignole, Francesca Borzumati, Sidney Coleman,
Savas Dimopoulos, Gia Dvali, Gian Giudice, Alexander Kusenko, and
Carlos Wagner.

\section*{Appendix}

We give here the threshold function $T(x)$. We recall first the
function $g(x)$ which determines gaugino masses:
\beq
g(x)={1 \over x^2}\bigl [(1+x)\ln(1+x)+(1-x)\ln(1-x)\bigr ]\,.
\label{eq:gdef}
\eeq
Consider a diagram with the topology of Fig.~\ref{fig:higgsino},
where the gaugino mass insertion is generated by a messenger loop.
Let the messenger sector be characterized by the following
parameters: $\Mess{}$ is the fermion mass, and $\Mess{}^2 (1\pm x)$
are the squared scalar masses. Then the gaugino mass $M$, or the
messenger loop at zero momentum, is given by
\beq
M = {\alpha\over 4\pi} \Lambda g(x)
\label{eq:gaugezer}
\eeq
where $\Lambda= x \Mess{}$. We now define a few useful quantities.
The first is the loop in Fig.~\ref{fig:higgsino}, evaluated with a
sharp infrared momentum cut off $k_{\rm IR}$. The result is $I_{\rm
UV}^G$, as defined in the text:
\beq
I_{\rm UV}^G = {\mu\over 8\pi^2} \int_{k_{\rm IR}}^\infty M(k) {dk
\over k} =
{\mu M\over 8\pi^2}\left \{\ln\left(\Mess{}/k_{\rm IR}\right) + 1 +
T(x) + {\cal O}(k_{\rm IR}^2/\Mess{}^2)\right \}
\label{eq:chi}
\eeq
where we indicate by $M(k)$ the gaugino mass at momentum $k$, so that
$M = M(0)$. We have set the vertex couplings for the external scalars
to 1. The threshold correction $T + 1$ implies that the gaugino mass
$M$, for RG purposes, is effectively turned on at a scale $\Mess{}
e^{1+T}$. The threshold function $T$ is given by
\bea
T(x) &=& -\frac12 + {1\over 4 x^2 g(x)} \bigl [(1+x)\ln^2(1+x)
+(1-x)\ln^2(1-x)    \nonumber \\
&{\phantom =}&\phantom{.} + 2 (1+x) \li2({x\over
1+x})+2(1-x)\li2({x\over x-1})-
2\li2(-x)-2\li2(x)
\bigr ]
\label{eq:Tfunc}
\eea
where $\li2$ is the dilogarithm function: $\li2(x) = -\int_0^1
[\ln(1-xt)/t] dt$. Note that $T(0) = 0$ and $T$ monotonically
decreases between $0$ and $1$, with $T(1) = \pi^2/(48 \log 2) - 1/2
\simeq -0.203$. However for less extreme $x$ choices away from unity,
$T$ is much smaller, and already at $x = 0.5$ its value is $\simeq
-0.026$. Another useful quantity is a loop like
Fig.~\ref{fig:higgsino}, evaluated at external Euclidean momentum
$k_E$ for the Higgs fields, with $k_E$ acting now as the infrared
regulator and $k_{\rm IR} = 0$. The result is obtained from the above
equation for $I_{\rm UV}^G$ just by changing the expression inside
brackets to
\beq
\ln(\Mess{}/k_E) + \frac32 + T(x) + {\cal O}(k_E^2/\Mess{}^2)\,.
\label{eq:newexpr}
\eeq
Notice, finally, that this result also yields $A(k)$ as defined in
the text. Indeed $A(k)$ is given by the same diagram with the Higgs
fields replaced by sfermions and with a Yukawa coupling replacing
$\mu$ on one of the
internal fermion lines.

Finally we give for completeness the expression of $L_{G1}$. Since
$M_1$ gets contribution from both the triplet and the doublet
messengers, we find:
\bea
L_{G1} &=& \eta(x_2,x_3) \left [\ln\left({\Mess3 \over \mu}\right) +
T(x_3) \right ] +
[1-\eta(x_2,x_3)] \left [\ln({\Mess2 \over \mu}) + T(x_2) \right ]
\nonumber\\
&{\phantom =} & \phantom{.} + 1 - {M_1^2 \over \mu^2 - M_1^2} \ln
\left({\mu\over M_1}\right)
\label{eq:Lone}
\eea
where $\eta(x_2,x_3) \equiv 2 g(x_3)/[2g(x_3)+3g(x_2)]$ measures the
relative contribution to $M_1$ from doublet and triplet messengers.


\begin{thebibliography}{99}

%
\bibitem{ref:old} M. Dine, W. Fischler and M. Srednicki, Nucl.\
Phys.\ {\bf B189}, 575 (1981); S. Dimopoulos and S. Raby, Nucl.\
Phys.\ {\bf B192} 353 (1981); L. Alvarez Gaum\`e, M. Claudson and M.
Wise, Nucl.\ Phys.\ {\bf B207} 96 (1982); C. Nappi and B. Ovrut,
Phys.\ Lett.\ B\ {\bf 113} 175 (1982).
%
\bibitem{ref:kostel} L.J. Hall, V.A. Kostelecky, and S. Raby, Nucl.\
Phys.\ {\bf B267} 415 (1986).
%
\bibitem{ref:kaplouis} V.S. Kaplunovsky and J. Louis, Phys.\ Lett.\ B
{\bf 306} 269 (1993).
%
\bibitem{ref:prog} K. Intriligator and N. Seiberg, hep-th/9509066,
proceedings of Trieste '95 spring school, TASI '95,
Trieste '95 summer school, and Carg\`ese '95 summer school.
%
\bibitem{ref:new} M. Dine and A. Nelson, Phys. Rev. D {\bf 48}, 1277
(1993); M. Dine, A. Nelson and Y. Shirman, Phys. Rev. D {\bf 51},
1362 (1995); M. Dine, A. Nelson, Y. Nir and Y. Shirman, Phys. Rev. D
{\bf 53}, 2658 (1996).
%
\bibitem{ref:phenom} D.R. Stump, M. Wiest and C.P. Yuan, Phys. Rev. D
{\bf 54}, 1936 (1996); S. Dimopoulos, M. Dine, S. Raby and S. Thomas,
Phys. Rev. Lett. {\bf 76}, 3494 (1996); S. Ambrosanio, G.L. Kane,
G.D. Kribs, S.P. Martin and S. Mrenna, {\it ibid.} 3498 (1996) and
Ref.~\cite{ref:ambro};  S. Dimopoulos, S. Thomas and J. Wells, Phys.
Rev. D {\bf 54}, 3283 (1996).
%
\bibitem{ref:babu} K.S. Babu, C. Kolda and F. Wilczek, Phys. Rev.
Lett. {\bf 77}, 3070 (1996).
%
\bibitem{ref:dnirs} M. Dine, Y. Nir and Y. Shirman, hep-ph/9607397.
%
\bibitem{ref:nelran} A.E. Nelson and L. Randall, Phys. Lett. B {\bf
316}, 516 (1993).
%
\bibitem{ref:hrs} L.J. Hall, R. Rattazzi and U. Sarid, Phys. Rev. D
{\bf 50}, 7048 (1994).
%
\bibitem{ref:us} R. Rattazzi and U. Sarid, Phys. Rev. D {\bf 53},
1553 (1996).
%
\bibitem{ref:martin} S. Dimopoulos, G.F. Giudice and A. Pomarol,
hep-ph/9607225; S.P. Martin, hep-ph/9608224.
%
\bibitem{ref:gdp} G.F. Giudice, G. Dvali and A. Pomarol, Nucl. Phys.
B {\bf 478}, 31 (1996).
%
\bibitem{ref:gjp} T. Gherghetta, G. Jungman and E. Poppitz,
hep-ph/9511317.
%
\bibitem{ref:higgsmass} J. Ellis, G. Ridolfi and F. Zwirner, Phys.
Lett. B {\bf 257}, 83 (1991); H. Haber and R. Hempfling, Phys. Rev.
Lett. {\bf 66}, 1815 (1991); Y. Okada, M. Yamaguchi, T. Yanagida,
Prog. Theor. Phys. {\bf 85}, 1 (1991); R. Barbieri, M. Frigeni and F.
Caravaglios, Phys. Lett. B {\bf 258}, 167 (1991).
%
\bibitem{ref:dtw} S. Dimopoulos, S. Thomas and J. Wells,
hep-ph/9609434.
%
\bibitem{ref:bg} R. Barbieri and G. Giudice, Nucl. Phys. B {\bf 306},
63 (1988); G.W. Anderson and D.J. Casta\~no, Phys. Lett. B {\bf 347},
300 (1995).
%
\bibitem{ref:ciastru} P. Ciafaloni and A. Strumia, hep-ph/9611204.
%
\bibitem{ref:mv} S.P. Martin and M.T. Vaughn, Phys. Rev. D {\bf 50},
2282 (1994).
%
\bibitem{ref:hemp} R. Hempfling, Phys. Rev. D {\bf 49}, 6168 (1994).
%
\bibitem{ref:toby} T. Falk, K.A. Olive, L. Roszkowski, A. Singh and
M. Srednicki, hep-ph/9611325.
%
\bibitem{ref:coleman} S. Coleman, Phys. Rev. D {\bf 15}, 2929 (1977);
C.G. Callan and S. Coleman, Phys. Rev. D {\bf 16}, 1762 (1977).
%
\bibitem{ref:lattice} A. Kusenko, Phys. Lett. B {\bf 358}, 51 (1995);
A. Kusenko, P. Langacker and G. Segre, hep-ph/9602414; I. Dasgupta,
hep-ph/9610403.
%
\bibitem{ref:ddr} I. Dasgupta, B.A. Dobrescu and L. Randall,
hep-ph/9607487.
%
\bibitem{ref:sidney} S. Coleman, Nucl. Phys. B {\bf 298}, 178 (1988);
and private communication.
%
\bibitem{ref:qb} R. Rattazzi and U. Sarid, in preparation.
%
\bibitem{ref:komatsu} H. Komatsu, Phys. Lett. B {\bf 215}, 323
(1988).
%
\bibitem{ref:papri} H. Pagels and J.R. Primack, Phys. Rev. Lett. {\bf
48}, 223 (1982).
%
\bibitem{ref:ambro} S. Ambrosanio, G.L. Kane, G.D. Kribs, S.P. Martin
and S. Mrenna, hep-ph/9605398.
%
\bibitem{ref:damien} J.A. Bagger, K. Matchev, D.R. Pierce and R.
Zhang, hep-ph/9609444.
%
\bibitem{ref:mormur} T. Moroi, H. Murayama and M. Yamaguchi, Phys.
Lett. B {\bf 303}, 289 (1993).
%
\bibitem{ref:strumia} A. Strumia,  hep-ph/9604417.
%
\bibitem{ref:bsg} R. Garisto and J.N. Ng, Phys. Lett. B {\bf 315},
372 (1993); M.A. Di\`az, Phys. Lett. B {\bf 322}, 207 (1994); F.M.
Borzumati, Z. Phys. C {\bf 63}, 291 (1994).
%
\bibitem{ref:Bur} A.J. Buras, M. Misiak, M. M\"unz and S. Pokorski,
Nucl. Phys. B {\bf 424}, 374 (1994).
%
\bibitem{ref:misiak} K. Chetyrkin, M. Misiak, M. Munz,
hep-ph/9612313.
%
\bibitem{ref:desh} N.G. Deshpande, B. Dutta and S. Oh,
hep-ph/9611443.
%
\bibitem{ref:greub} C. Greub and T. Hurth, hep-ph/9608449.
%
\bibitem{ref:CLEO} M.S. Alam et al., Phys. Rev. Lett. {\bf 74}, 2885
(1995).
%
\bibitem{ref:GMSBYU} J.A. Bagger, K. Matchev, D.R. Pierce and R.
Zhang, hep-ph/9611229.


\end{thebibliography}
\end{document}